\documentclass[epj]{svjour}
\usepackage{graphics}
\usepackage{graphicx}
\usepackage{amssymb}
\usepackage{amsmath}
\usepackage{xspace} 

\usepackage{color}
\begin{document}
\definecolor{dg}{rgb}{0.00, 0.70, 0.40}
\newcommand{\nag}{\phantom{\dag}}
\newcommand{\ii}{\mathrm{i}}
\newcommand{\ee}{\mathrm{e}}
\newcommand{\hh}{\mathrm{h}}
\newcommand{\hf}[1]{\textcolor{red}{#1}}
\title{Heating and thermoelectric transport in a molecular junction}
\author{ Jan Loos\inst{1} \and Thomas Koch\inst{2} \and Holger Fehske\inst{2}}
\institute{Institute of Physics, Academy of Sciences of the Czech Republic, 16200 Prague, Czech Republic 
\and Institute of Physics,   Ernst Moritz Arndt University Greifswald, 17487 Greifswald, Germany
}
\date{Received: date / Revised version: date}

\abstract{
The energy dissipation and heat flows associated with the particle current
in a system with a molecular junction are considered. In this connection,
we determine the effective temperature of the molecular oscillator that is
compatible with the existence of a steady state. The calculations based on
the Kadanov-Baym nonequilibrium Green function formalism are carried out
supposing a strong coupling of the dot electrons with the molecular vibrations.
Accordingly, the representation given by the Lang-Firsov polaron
transformation is used and the dependence of results on the electron-phonon
interaction strength is investigated.
}
\maketitle
\section{Introduction}\label{SECintro}
The field of molecular electronics opens new and fascinating possibilities for analyzing how heat is carried, distributed, stored, and converted in nanoscale systems~\cite{Dubi2011,BR13}. These are very fundamental and timely questions not only from a theoretical point of view: The control of heat flows in electronic devices is one of the present day technological challenges with great practical impact on society~\cite{USDOE}. Of particular importance in this respect is an in-depth analysis  of thermoelectric phenomena in molecular junctions~\cite{EIA10}.  These systems are efficient thermoelectric devices, which can be engineered using synthetic chemistry~\cite{RJSM07}. For example, they can be used for reliable single-molecule measurements of barrier tunneling, the transmission function, the differential conductance, thermal currents or the thermopower.

The perhaps simplest theoretical model in this context is a two- or three-terminal setup, where the molecular bridge 
between macroscopic leads [having fixed (and possibly different) temperatures and chemical potentials] is replaced by a single resonant level. Then, in the Coulomb blockade regime, when the effects of spins and electronic correlations can be neglected, a transport electron residing on the dot level will only interact with the vibrational modes of the molecule~\cite{Galperin2007a}. The population of these phonons will be determined by this electron (in the sense of ``floating molecule") or by a coupling to the phonon source kept at a constant temperature as well~\cite{EIA10}.  The heating of the molecular junction by the energy dissipation is crucial for the thermal stability of the device~\cite{Scea08,RPD07}. To obtain an intuitive measurement quantity for the junction heating, an effective molecular (phonon) temperature is
introduced. This task is most simply accomplished by equating the steady-state vibronic energy with the equilibrium energy given by the Bose-Einstein phonon population. However, the analysis in~\cite{Galperin2007b} shows that this approach appears inconsistent because it gives incorrect results at low bias voltages in the case of  strong electron-phonon (EP) interaction. A correct determination of the effective temperature can be achieved by introducing an auxiliary phonon bath coupled to the molecular oscillator as a ``thermometer"~\cite{Galperin2007b}.

The electronic heat transport in such open quantum system is frequently studied by means of Green's function approaches which, following~\cite{BR13}, may be roughly divided into many-body perturbation theories, such as those based on approximations to Hedin's equations~\cite{He65}, the molecular Dyson equation~\cite{BS09}, or the Kadanoff-Baym equations~\cite{Kadanoff1962,DL07,MSSL08}, and effective single-particle methods based, e.g., on the Kohn-Sham scheme of density functional theory~\cite{TGW01,DGD02,CFR05}. 

In previous work~~\cite{LKABF09,Koch2010,Koch2011,Koch2012,Koch2014}, a non-linear response theory for charge transport through a molecular junction was developed by the authors.  
Since the particle current is accompanied by the heat flow between
the system and the environment~\cite{AM76,Mah00}, according to the first law of 
thermodynamics, the heat transfer has to be considered together with the work
needed for electrons to overcome the potential difference sitting on the
junction. In the present treatment, the energy dissipation and heat flows
in the steady state of the system will be consider for various boundary
conditions if a strong interaction between the dot electrons and the dot
molecular oscillator is supposed. Using the steady-state condition, the
effective temperature determining the oscillator populations in the 
nonequilibrium steady state will be self-consistently determined  within the theoretical framework~\cite{Koch2011,Koch2014}, which is based  on a Lang-Firsov (small polaron) transformation~\cite{LF62r} and the solution of the equations of 
motion in the Kadanoff-Baym Green-function formalism~\cite{Kadanoff1962}.

The paper is organized as follows. Section~2.1 specifies the molecular junction model under investigation  and our theoretical approach. Section~2.2 presents the relations for the  
energy dissipation and heat flows, and introduces the condition for the effective phonon temperature.  A detailed derivation of the formulas of Sec.~2 can be found in the Appendices~A, B, and C. The numerical results presented in Sec.~3 demonstrate the various thermoelectric effects, including their dependence on the EP coupling strength.  Sec.~4 provides our conclusions. 

\section{Theoretical approach}\label{SECintro}
\subsection{Model and Green function formalism}\label{SECmgf}
\begin{figure}[b]
\includegraphics[width=0.8\linewidth]{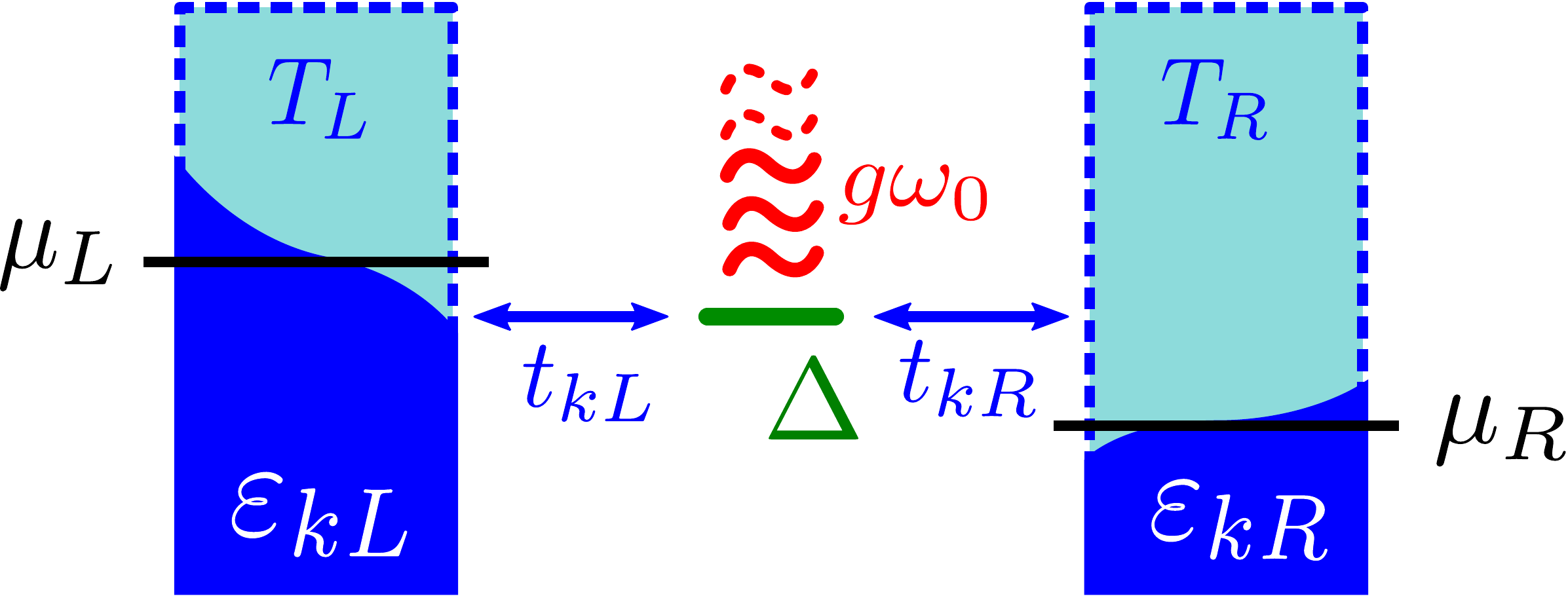}
\caption{Molecular quantum dot model under consideration. }
\label{FIGmodel}
\end{figure}
The molecular-junction being examined in a nonequilibrium steady state  is sketched in Fig.~\ref{FIGmodel}. Thereby the left and right macroscopic leads, labelled by  $a = L, R$, are supposed to  be in local equilibrium states, however,  and are characterized by the chemical potentials $\mu_{L,R}$ and temperatures $T_{L,R}$ kept constant  by contact to heat baths.  

We model this system by the Anderson-Holstein Hamiltonian
\begin{align}
\label{model}
 H  &= \Delta \,d^\dag d^{\nag} -\; g \omega_0 \,d^\dag d \,( b^{\dag} + b) + \omega_0 \,b^{\dag} b \\
&+ \sum_{k,a} \left(\varepsilon_{ka}-\mu_a\right) c_{ka}^{\dag} c_{ka}^{\nag} - \frac{1}{\sqrt{N}}\sum_{k, a}\big(t_{ka}^{\nag}d^{\dag}c_{ka}^{\nag} + t_{ka}^\ast c_{ka}^{\dag}d\big).\nonumber
\end{align}
Here, the electronic dot degrees of freedom, represented by the fermionic operators  $d^{(\dagger)}$, are coupled locally, with an interaction strength $\propto g$, to the vibrations of the molecule of  frequency $\omega_0$, taken into account  by the bosonic operators $b^{(\dagger)}$. The last term on the r.h.s. of~\eqref{model} gives the electron hopping between the quantum dot level $\Delta$ and the macroscopic leads. The leads are described by two sets of free fermion states (created by $c_{ka}^{\dag}$), having energies  $\varepsilon_{ka}$ and chemical potentials $\mu_a$. In the definition of energies we put the equilibrium chemical potential $\mu_{eq}=0$, neglecting the temperature dependence of the Fermi levels  in the leads for the temperature changes considered.

Looking for the response of the system to a potential difference between the leads, we assume that the potentials are realized by a symmetric voltage drop across the junction, 
\begin{align}
\mu_L&=-\mathrm{e}\,\frac{\Phi}{2}\,,\quad \mu_R=\mathrm{e}\,\frac{\Phi}{2}\label{EQUmus}
\end{align}
with $\Phi >0$, where the electron charge $\mathrm{e}<0$. Since we assumed the lead Fermi level of the unperturbed system to  be zero, the part of the Hamiltionian~\eqref{model} corresponding to the disturbance by the bias voltage can be written as
\begin{align}
H_{U} &=\sum_{a}U_a\sum_{k}c_{ka}^\dag c_{ka}^{\nag}\,,\quad U_a=-\mu_a \,.
\label{disturbance}
\end{align}
According to the formalism developed in~\cite{Kadanoff1962},  this interaction Hamiltonian is used below to determine the non-linear response of the system.

The main subject of the subsequent considerations is the steady-state electron transport through the molecular junction for given boundary conditions. Owing to particle conservation, the particle current out of lead $a$ is given by 
 \begin{align}
\hat J_a
&=\frac{\mathrm{i}}{\sqrt{N}} \sum_k \Big[ t_{ka} d^\dag c_{ka}^{\nag} - t_{ka}^\ast c_{ka}^\dag d \Big]\,.\label{EQUcurrentop}
\end{align}
The mean values of the operators $\hat J_a$ will be determined by the nonequilibrium  real-time Green functions of dot operators $d^{(\dagger)}$ and lead operators $c_{ka}^{(\dag)}$, defined as in~\cite{Kadanoff1962}. For example, we have 
\begin{align}
g_{dd}^<(t_1,t_2;U)&=\;\;\;\mathrm{i}\langle  d_U^\dagger(t_2)d_U(t_1)\rangle \,,\label{EQUdefresponseless}\\
g^>_{dd}(t_1,t_2;U)&=-\mathrm{i}\langle  d_U(t_1)d_U^\dagger(t_2)\rangle \label{EQUdefresponsegtr}\,.
\end{align}
Here, the time dependence of the operators in the Heisenberg picture is given by the full operator $H$, whereas the mean value $\langle \cdots \rangle$ stands for the statistical average  before the disturbance~\eqref{disturbance} was turned on. In particular, calculating the steady-state mean value of $\hat J_a$ the relation
\begin{align}
\mathrm{i}\langle d^\dag c_{ka}^{\nag} \rangle &= g_{cd}^<(k,a;t_1,t_1;U)
\end{align}
is used, and a similar one for  $\langle c_{ka}^{\dag} d \rangle$. In this way, the calculations detailed  in~\cite{Koch2011} lead to the following result for $J_a=\langle \hat J_a\rangle$:
\begin{align}
J_a&= \int_{-\infty}^{\infty}\frac{\mathrm{d}\omega}{2\pi}\; \Gamma_a^{(0)}(\omega)\Big \{ f_a(\omega+U_a)\, g_{dd}^{>}(\omega;U) \nonumber\\
& \qquad \qquad-\left[ 1-f_a(\omega+U_a)\right]\,g_{dd}^{<}(\omega;U)  \Big \}\,, \label{EQUcurrentJa}
\end{align}
where 
\begin{align}
f_a(\omega)&=(e^{\beta_{a}\omega}+1)^{-1}\quad\mbox{with}\quad \beta_{a}=(k_B T_a)^{-1}\label{EQUfermileads}
\end{align}
is the lead Fermi function and
\begin{align}
\Gamma^{(0)}_a(\omega)&=2\pi \frac{1}{N}\sum_k  |t_{ka}|^2 \delta(\omega-\varepsilon_{ka})
\end{align}
characterizes the rate of the dot-lead hopping.  Assuming identical leads the index $a$ at $\Gamma^{(0)}_a$ can be omitted.

In the steady state $J_R=-J_L$ the particle current through the junction may be expressed with the use of the electronic spectral function $A(\omega,U)$ as 
\begin{align}
J&=\frac{1}{2}(J_L-J_R)\label{EQUcurrent}
=\int_{-\infty}^{\infty}\frac{\mathrm{d}\omega}{4\pi}\; \Gamma^{(0)}(\omega)A(\omega;U)\\&\hspace*{3.4cm}\times \left [ f_L(\omega+U_L) -f_R(\omega+U_R)  \right ].\nonumber
\end{align}
Taking $t=t_1-t_2$, the Fourier transforms of $g^{\lessgtr}_{dd}$, defined according to~\cite{Kadanoff1962} as 
\begin{align}\label{ftg}
g_{dd}^{\lessgtr} (\omega;U) &= \mp\mathrm{i}\int_{-\infty}^{\infty} \mathrm{d}t\;g_{dd}^{\lessgtr}(t;U)e^{\mathrm{i}\omega t}\,,
\end{align}
determine the electron spectral function 
\begin{align}
A(\omega,U)=g^>_{dd}(\omega,U)+g^<_{dd}(\omega,U)\,.
\end{align}
As a consequence, the non-linear response of our system to the bias voltage and/or the temperature difference between the leads is determined by the nonequilibrium electron Green functions and the electron spectral function. 

For a strong coupling of the dot electrons to the molecular vibrations, the bare-electron states are not  a suitable basis for the calculation. Here, the situation resembles the Holstein polaron problem if the EP interaction is strong in comparison to the (bare) hopping kinetic energy. That is why the polaron representation is used in Appendix~\ref{AA} and the electronic Green functions are calculated by means of the polaronic ones. Henceforth, the electronic operators and functions will be denoted by tilde symbols, whereas symbols without tilde belong to the polaronic operators/functions.

\subsection{Dissipation and heat flow}\label{SECdhf}
The theoretical background outlined in the preceding section, together with the explicit expressions of the Green functions given in Appendix~\ref{AB}, will now be applied to specific situations defined by different types of boundary conditions. In doing so, the dissipation and heat flows needed to maintain the steady-state regime will be discussed.

The steady-state conditions will be formulated on the macroscopic level for the whole system (leads plus dot) if the leads are connected to the heat baths of temperatures $T_L$ and $T_R$, respectively, and the battery causes the potential difference  $\Phi$ across the junction.  

On account of the first law of  thermodynamics,
\begin{align}
dE=\delta Q+dW\,,
\label{firstlaw}
\end{align}
we equate the increase in energy of the whole system, $dE$, with the input of energy in the system by heat transfer, $\delta Q$, and the work $dW$ performed by the environment on the system. The work needed for $Jdt$ electrons to cross from $L$ to $R$ and overcome the potential difference sitting on the molecular junction is equal to
\begin{align}
\mathrm{d}W&=-\mathrm{e}\,\Phi J \mathrm{d}t\,.\label{EQUdW}
\end{align} 
The heat flow $J_a^Q$ into the lead $a$ from the heat bath at temperature $T_a$ is given by $(-dE_a/dt)$, where  $E_a$ is the sum of the lead-electron energies measured from the chemical potential $\mu_a$~\cite{AM76,Mah00}. The operator corresponding to $E_a$ is represented by the fourth term in~\eqref{model}. In a similar way as for the particle current $J_a$, the following result for $J_a^Q$ is obtained:
\begin{align}\label{EQUcurrentJQa}
J_a^Q&= \int_{-\infty}^{\infty}\frac{\mathrm{d}\omega}{2\pi}\; \omega\;\Gamma^{(0)}(\omega)\Big \{ f_a(\omega+U_a)\widetilde g^{>}_{dd}(\omega;U) \\
&\hspace*{1.8cm}-\left[ 1-f_a(\omega+U_a)\right]\widetilde g^{<}_{dd}(\omega;U)  \Big \}-\mu_aJ_a\,.\nonumber
\end{align}  
Let us stress again, that now the $\widetilde g^{\lessgtr}_{dd} (\omega;U)$ denote  the electronic Green functions---defined by Eqs.~\eqref{EQUdefresponseless}, \eqref{EQUdefresponsegtr}, and \eqref{ftg}---in the polaronic representation. In this representation the electronic operators are  given  by~\eqref{EQUtilded}, and the tilde Green functions are expressed by~\eqref{EQUgelec}  in terms of the polaronic Green functions.  The relation between electronic and polaronic Green functions was derived in~\cite{Koch2011} (see Sect. F), where also decoupling of the polaronic and shifted oscillator averaging was discussed in great detail.

As a result, the contribution of the leads to $\delta Q$ on the r.h.s.~of~\eqref{firstlaw} is given by $(J_L^Q+J_R^Q)dt$. In general, also the phonon bath, to which the molecular oscillator is coupled, has to be considered. In fact, the calculations recapitulated in Appendix~\ref{AC}, assuming a fixed phonon temperature $T_P$, correspond to the case of a strong coupling of the molecular oscillators to the phonon heat bath. In this case, the steady state condition $dE=0$ gives the heat flow from the phonon heat bath into the system as
\begin{align}
J_P^Q&=-(J_L^Q+J_R^Q)+\mathrm{e}\Phi J \, .
\label{jpq}
\end{align}
Here, the $J^Q_a$ are obtained by fixed model parameters, including the phonon heat-bath temperature $T_P$ that determines the populations of the local oscillator by the Bose-Einstein distribution
\begin{align}
n_B(\omega)&=(e^{\beta_P\omega}-1)^{-1}\quad\mbox{with}\quad\beta_P=(k_BT_P)^{-1}\,.
\label{boseeinstein}
\end{align}

The opposite limiting situation is represented by a negligible direct coupling of the molecular oscillator to the phonon bath. Then  the oscillator populations are determined mostly by the EP interaction. If we accept the theoretical approach formalized in the Appendix~\ref{AC}, the temperature $T_P$ entering there has to be considered as an effective temperature $T_P^{eff}$ that  is compatible with the existence of the steady state. Putting $dE=0$, i.e., $J_L^Q+J_R^Q-\mathrm{e}\Phi J =0$, the equation for the dissipation of the work into the heat taken away through the electron channel explicitly reads
\begin{align}\label{EQUbilanz}
 0=&\int_{-\infty}^{\infty}\frac{\mathrm{d}\omega}{2\pi}\; \omega\sum_a\Gamma^{(0)}(\omega)\Big \{ f_a(\omega+U_a)\widetilde g_{dd}^{>}(\omega;U) \\
&-\left[ 1-f_a(\omega+U_a)\right]\widetilde g_{dd}^{<}(\omega;U)  \Big \}-\sum_a\mu_aJ_a -\mathrm{e}\Phi J\nonumber\,,
\end{align}
where for the stationary particle current we have $-\sum_a\mu_aJ_a=\tfrac{1}{2}\mathrm{e}\Phi (J_L-J_R)=\mathrm{e}\Phi J$.  Inserting~\eqref{EQUgelec} for $\widetilde g_{dd}^{\lessgtr} (\omega;U)$ into~\eqref{EQUbilanz}, the condition for the determination of the effective temperature 
$T_P^{eff}$ becomes
\begin{align}\label{EQUdiss}
0&=\int_{-\infty}^{\infty}\frac{\mathrm{d}\omega}{2\pi}\sum_{s\ge 1} I_{s}(\kappa)2\sinh(s\theta)\,s\omega_0 \, \mathrm{e}^{-g^2\coth\theta} \nonumber\\
&\quad\times\sum_a\Big\{ \Gamma_a^{(0)}(\omega+s\omega_0) \nonumber\\
&\quad\times\Big(\big[1+n_B(s\omega_0)\big]g^{>}_{dd}(\omega;U)f_a(\omega+s\omega_0+U_a) \nonumber\\
&\quad- n_B(s\omega_0)g^{<}_{dd}(\omega;U)\big[1-f_a(\omega+s\omega_0+U_a)\big] \Big )\nonumber\\
&\quad + \Gamma_a^{(0)}(\omega-s\omega_0) \nonumber\\
&\quad\times \Big ( \big[1+n_B(s\omega_0)\big]g^{<}_{dd}(\omega;U)\big[1-f_a(\omega-s\omega_0+U_a)\big]\nonumber\\
&\quad- n_B(s\omega_0) g^{>}_{dd}(\omega;U)f_a(\omega-s\omega_0+U_a) \Big ) \Big\}\,.
\end{align}
The latter equation has the form of a balance equation for the phonon-assisted transitions in which an electron passes from an occupied state of the leads to the unoccupied polaronic dot-state, and vice versa.

\section{Numerical results and Discussion}
\begin{figure}[t]
\includegraphics[width=0.44\linewidth]{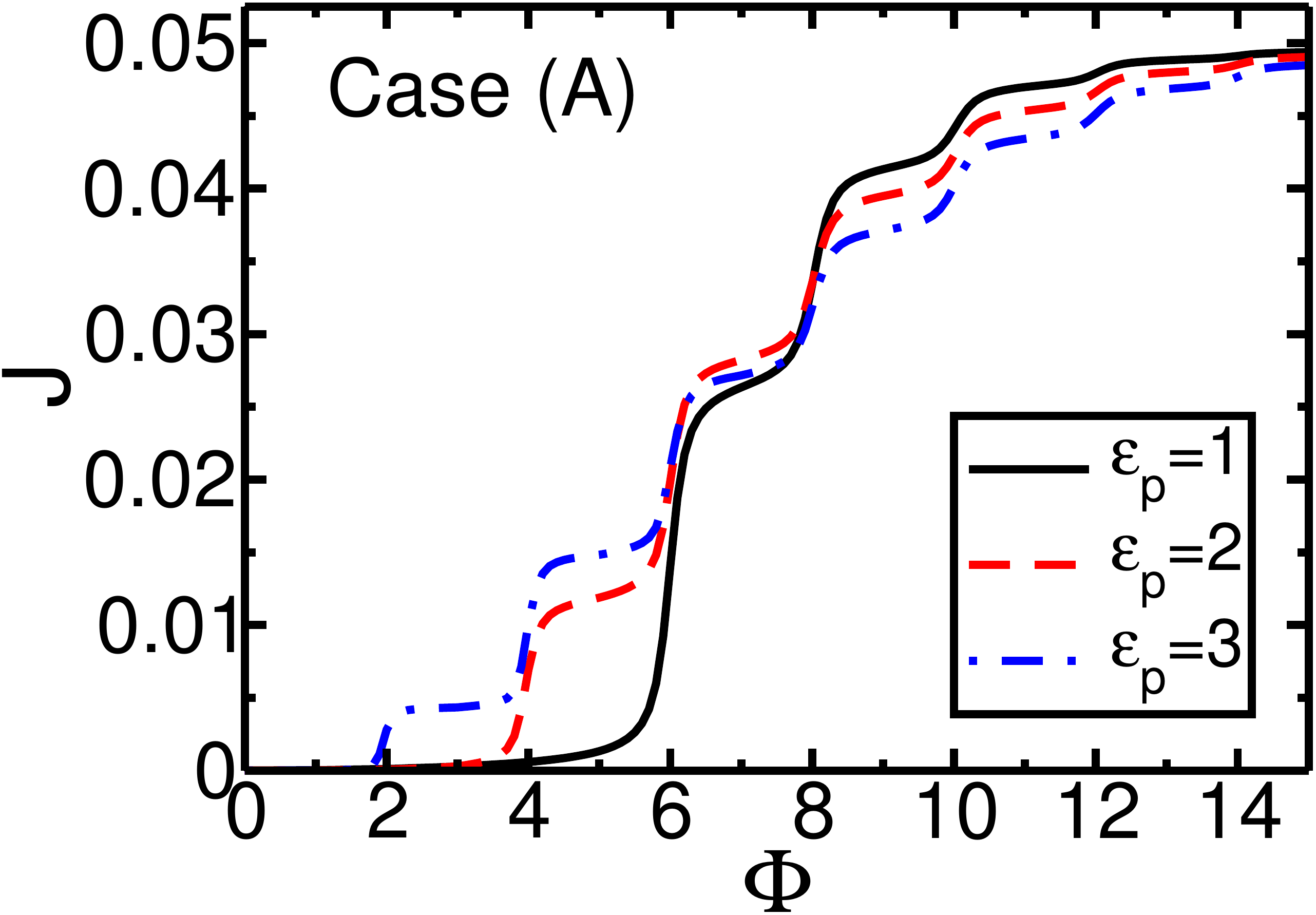}$\;$\includegraphics[width=0.44\linewidth]{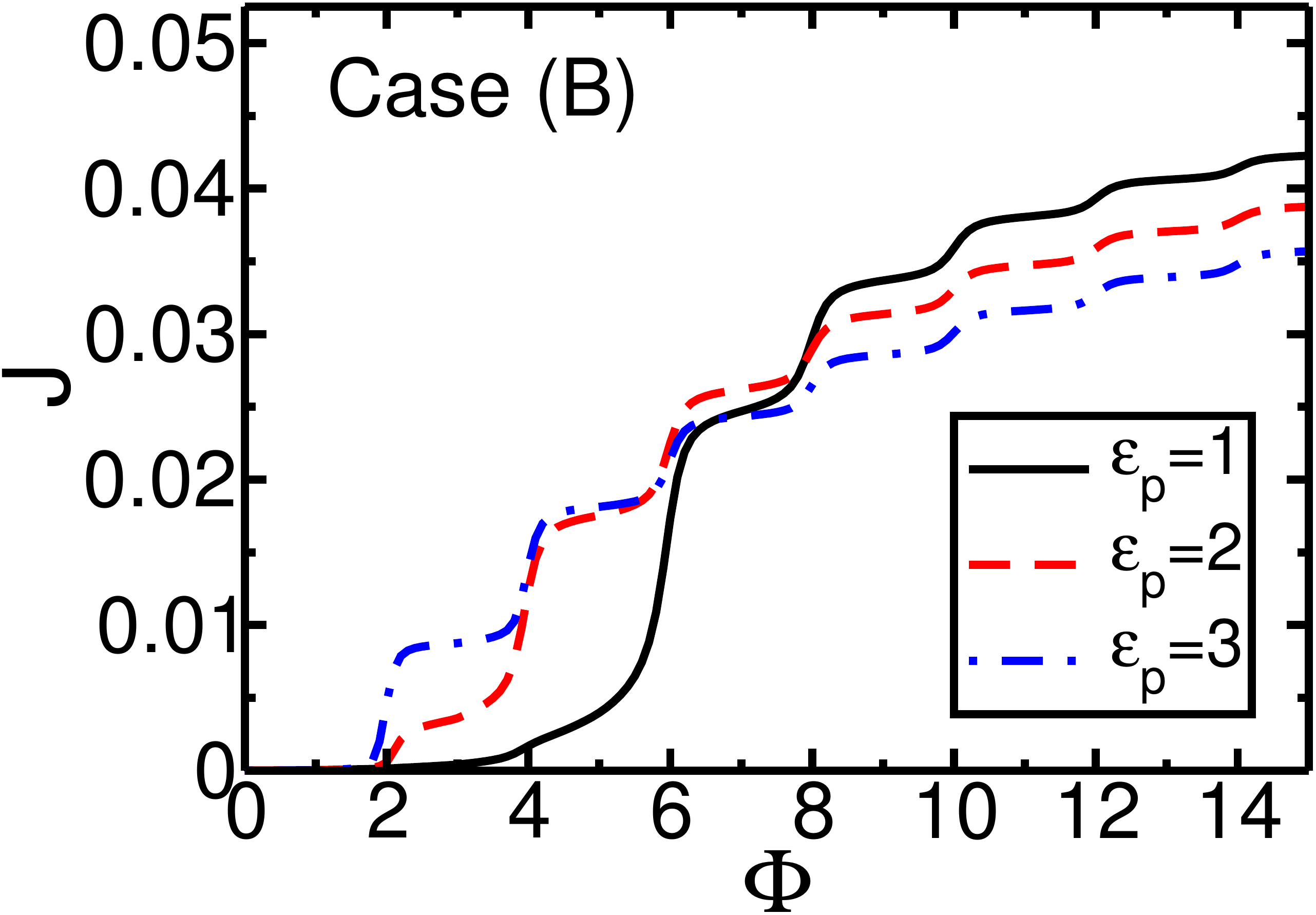}\\[0.3cm]
\includegraphics[width=0.44\linewidth]{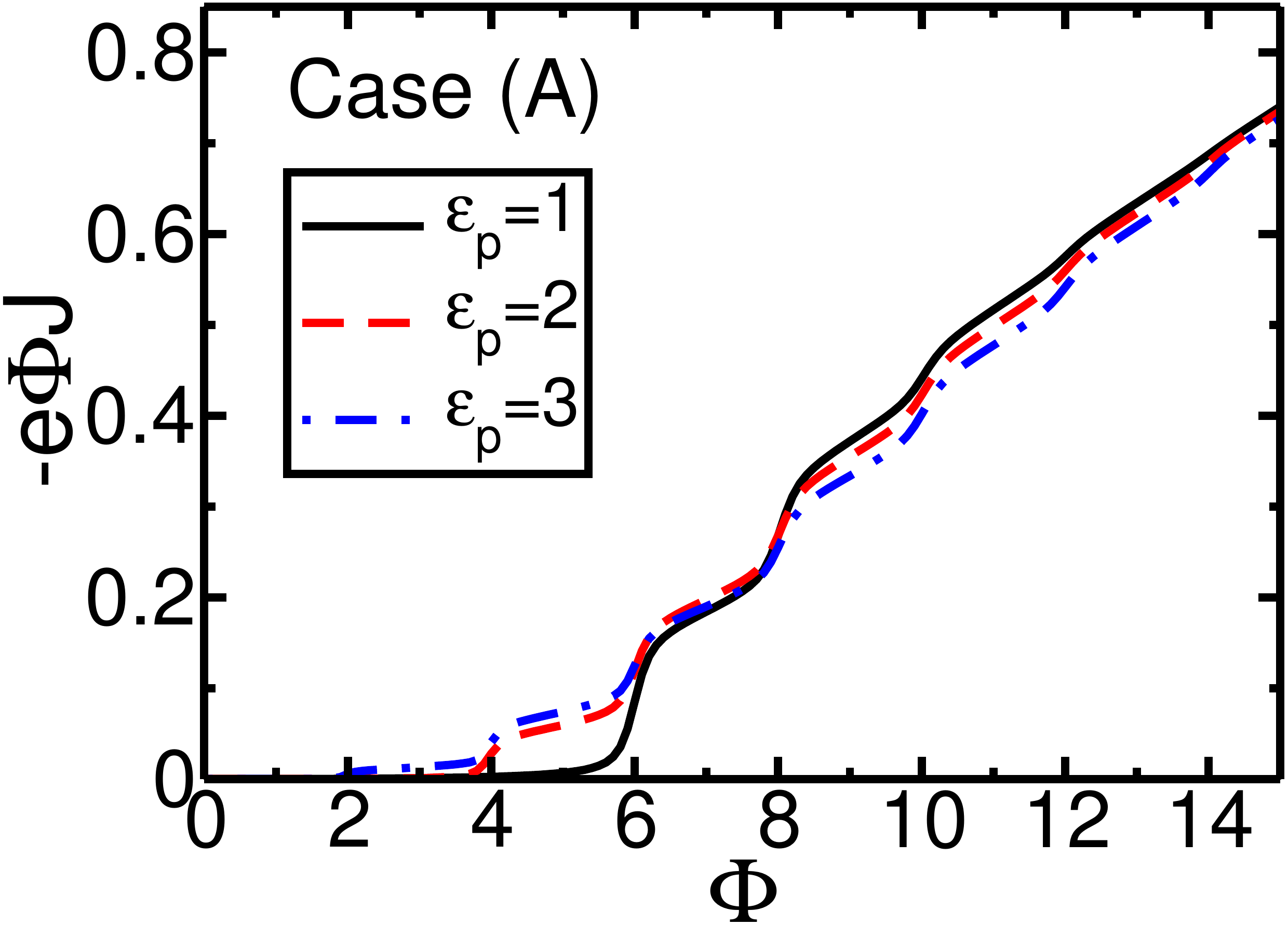}$\;$\includegraphics[width=0.44\linewidth]{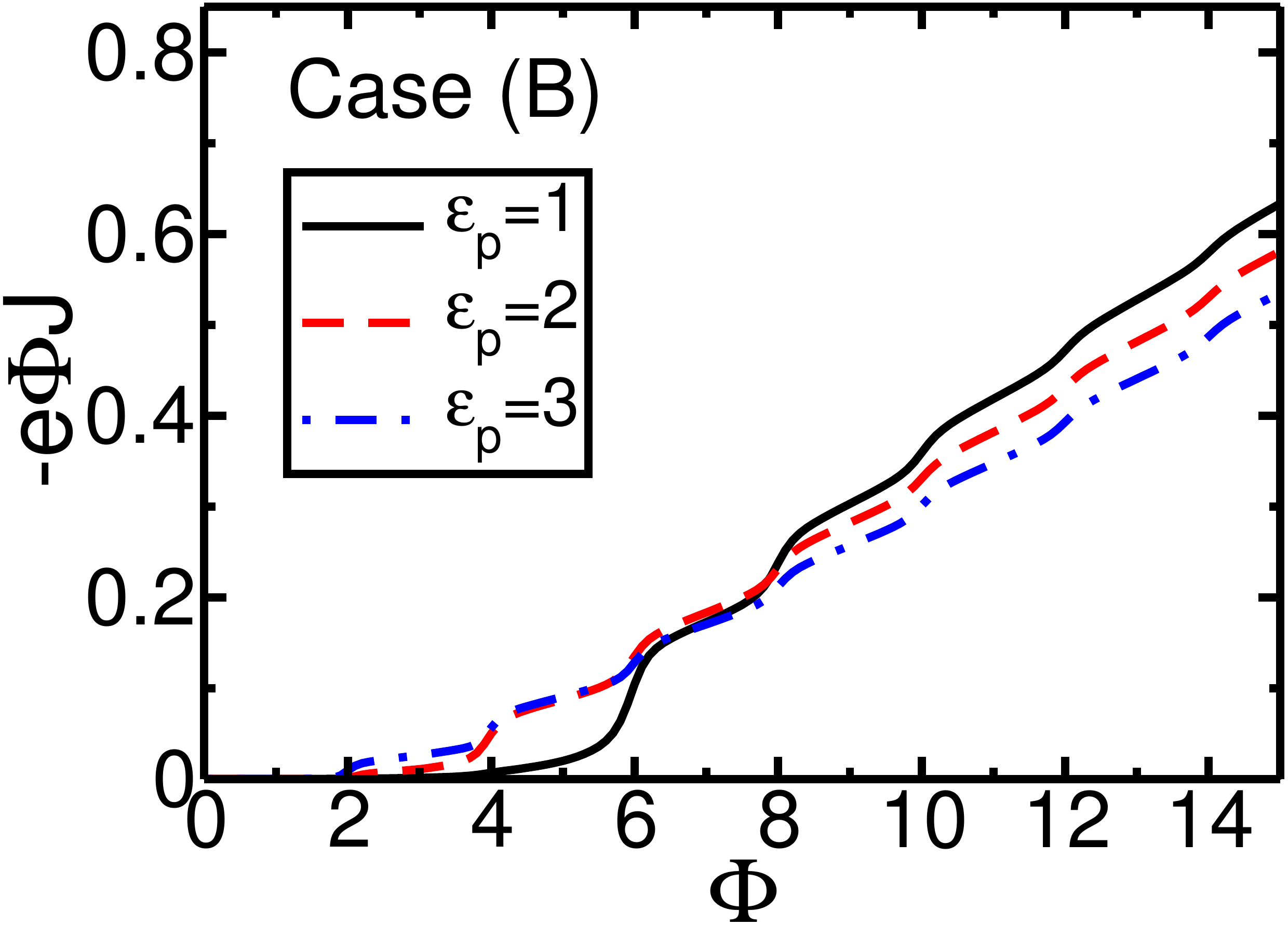}\\[0.3cm]
\includegraphics[width=0.46\linewidth]{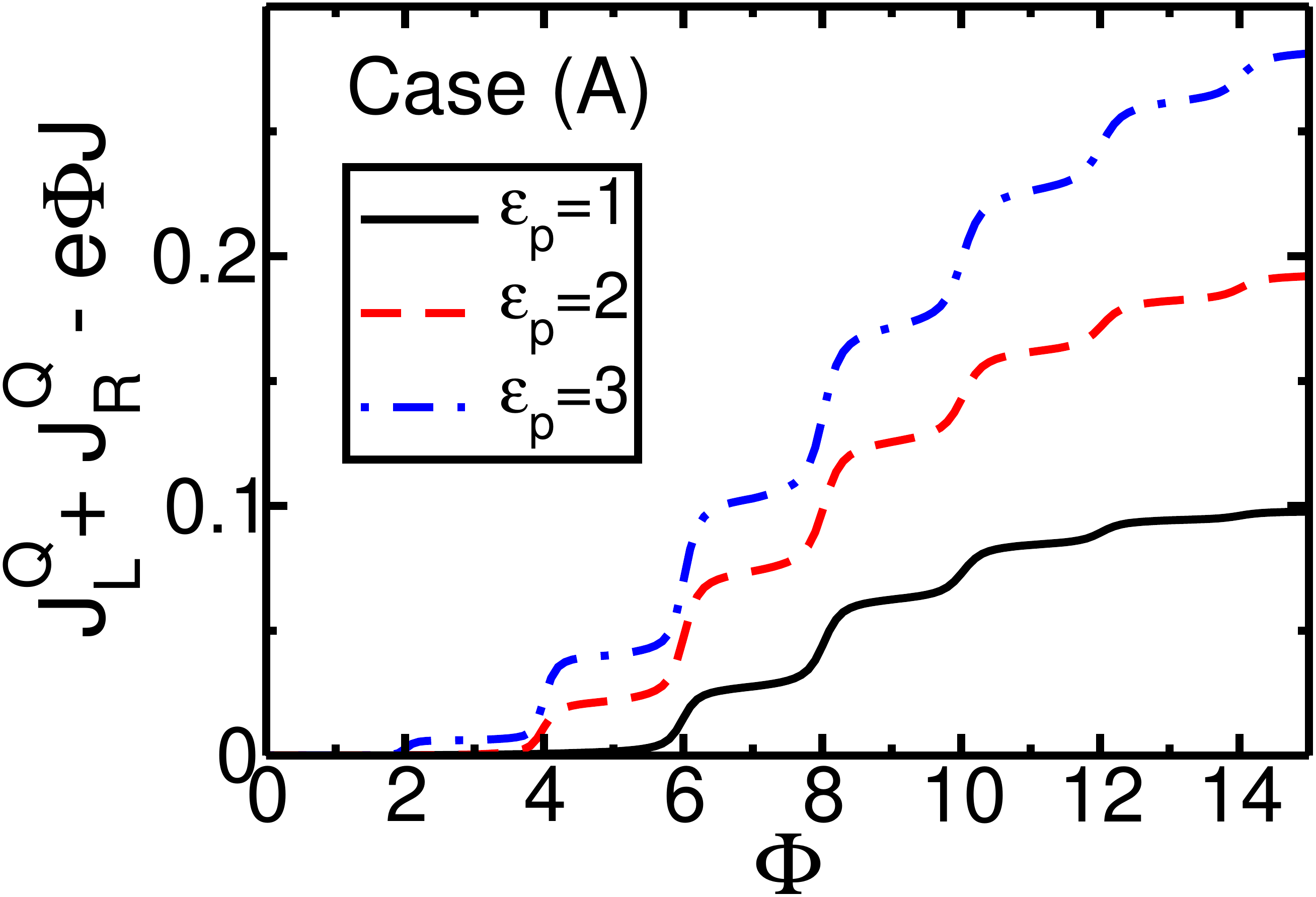}$\;$\includegraphics[width=0.44\linewidth]{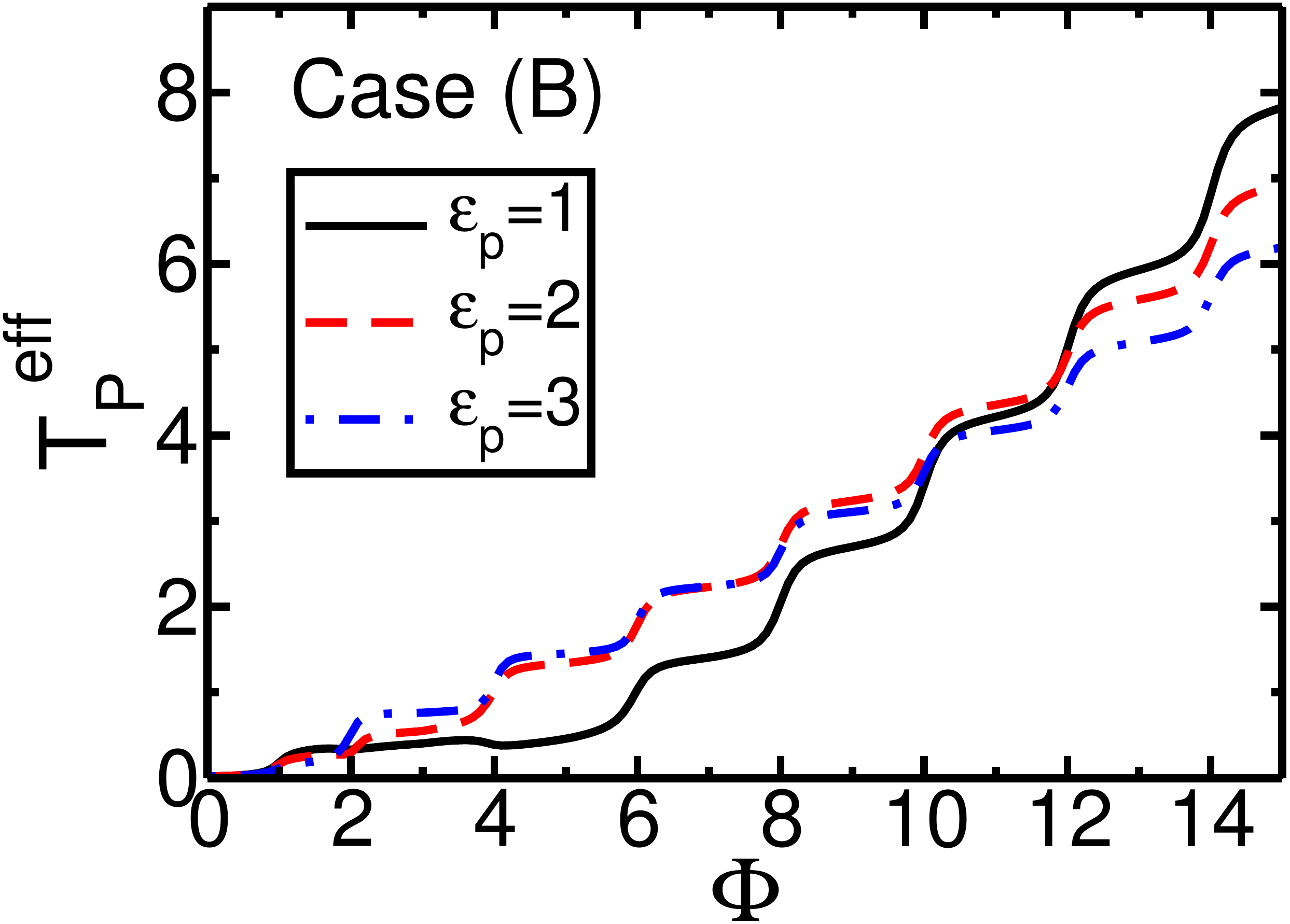}\\[0.3cm]
\includegraphics[width=0.46\linewidth]{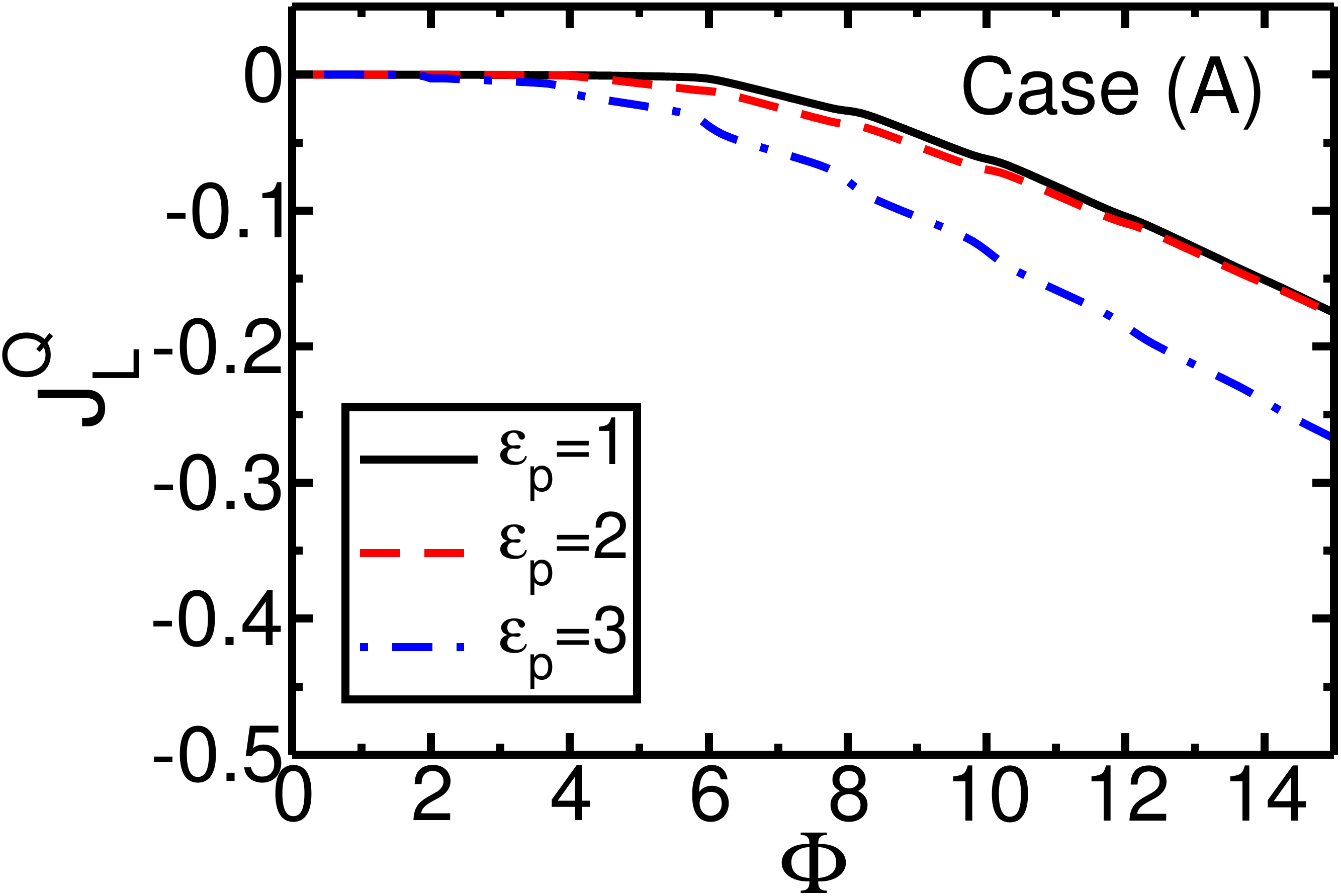}$\;$\includegraphics[width=0.46\linewidth]{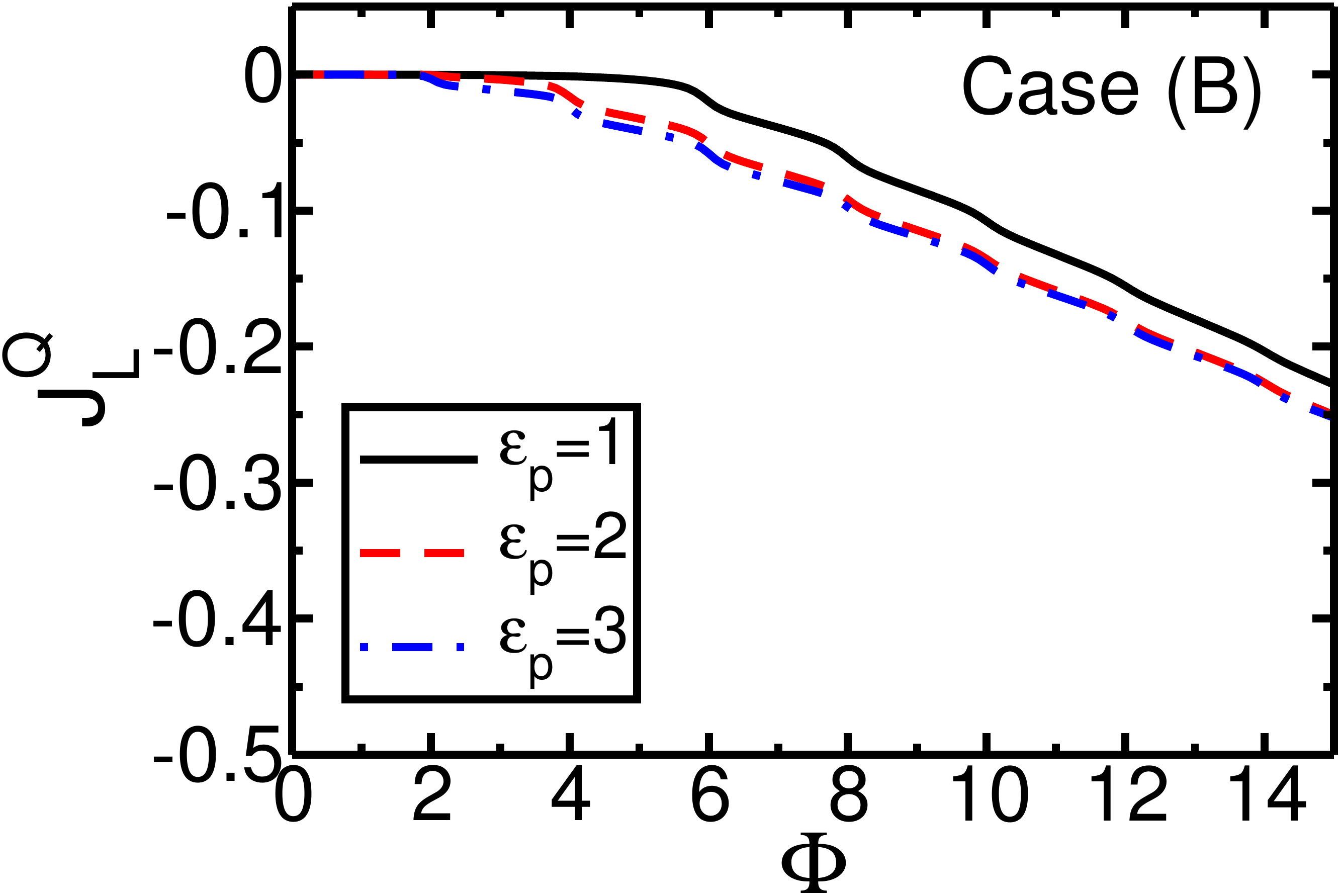}\\[0.3cm]
\includegraphics[width=0.46\linewidth]{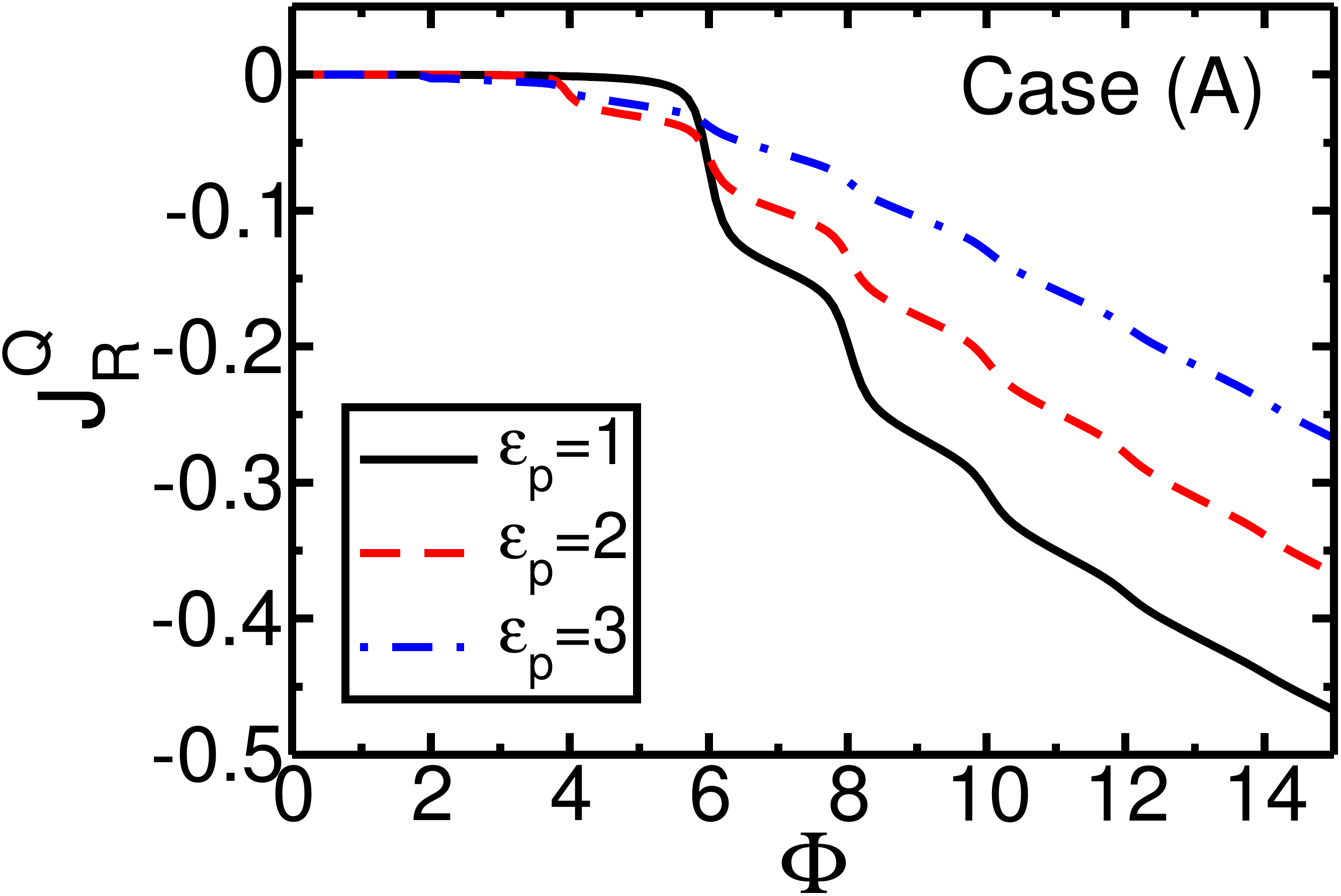}$\;$\includegraphics[width=0.46\linewidth]{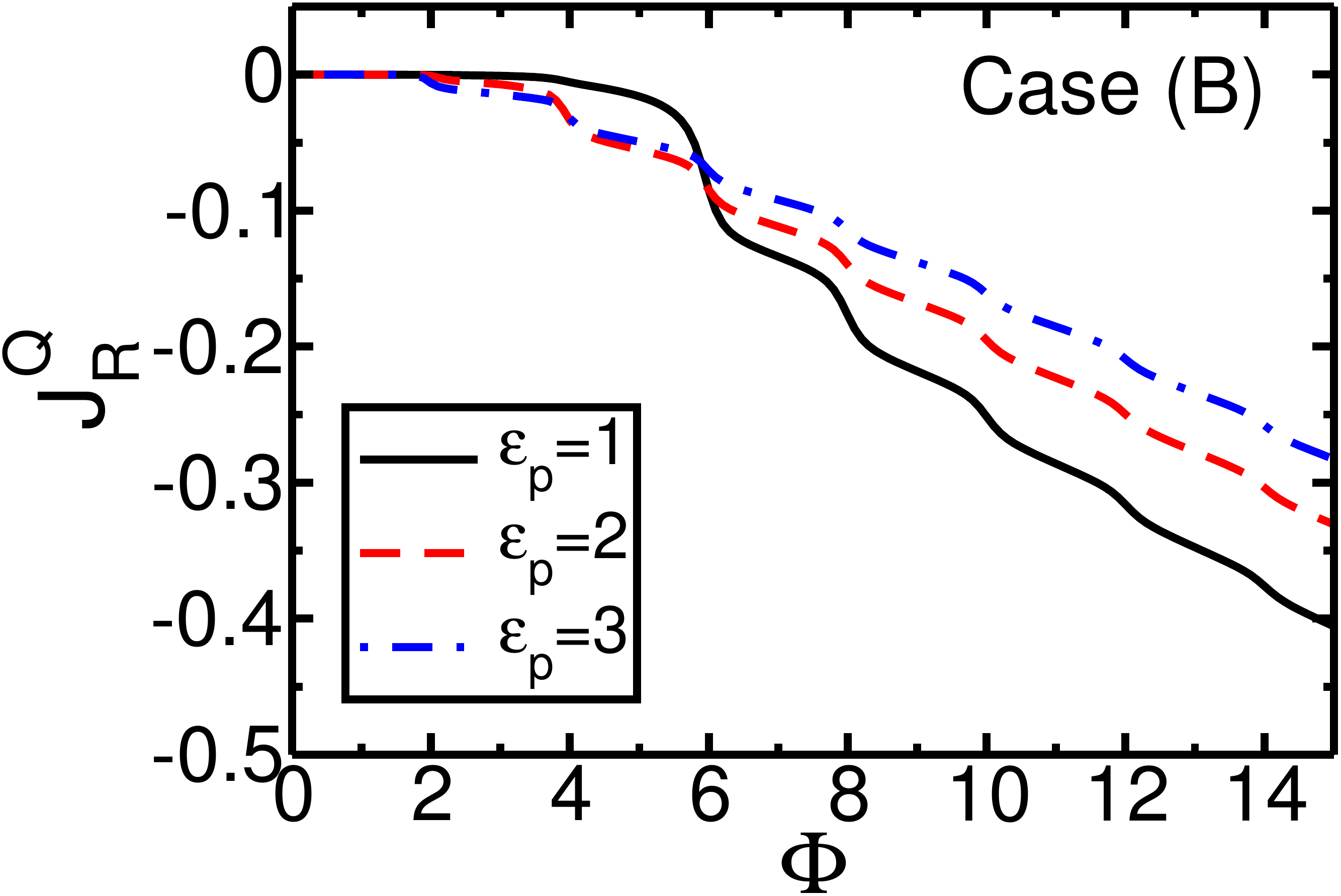}
\caption{Characteristics of thermoelectric transport through a molecular junction for fixed $T_P$ [case (A), left panels] and a self-consistently determined (effective)  temperature $T_P^{eff}$ [case (B), right panels], where $\Gamma_0=0.1$, $\Delta=4$, $T_R=0.02$,  and $\delta T=0$.  First row: Current $J$ as a function of the voltage bias $\Phi$. Second row: Dissipation at the junction. Third row: r.h.s.~of Eq.~(\ref{EQUdiss}) (left) and effective temperature determined from the steady-state condition (right). 
Fourth  and fifth row: Heat currents from the leads to the dot.}
\label{FIG2}
\end{figure}
In this section, the above reasonings  will be corroborated by numerical calculations performed for the both mentioned cases: A fixed $T_P$ [denoted in what follows by (A)] and  a self-consistently determined $T_P^{eff}$ [denoted by (B)]. 
Then, in case (A), $T_P=T_R$ and $T_L=T_R+\delta T$, whereas, in case (B), $T_P^{eff}(\Phi)$ results from the solution of~\eqref{EQUdiss} because the $U_a$ are given by~\eqref{EQUmus},~\eqref{disturbance}.   
In what follows, we take the bare phonon frequency as the unit of energy, i.e., we keep $\omega_0=1$ fixed, and set $\hbar=1$, $|\mathrm{e}|=1$, and $k_B=1$.

\begin{figure}[t]
\includegraphics[width=0.44\linewidth]{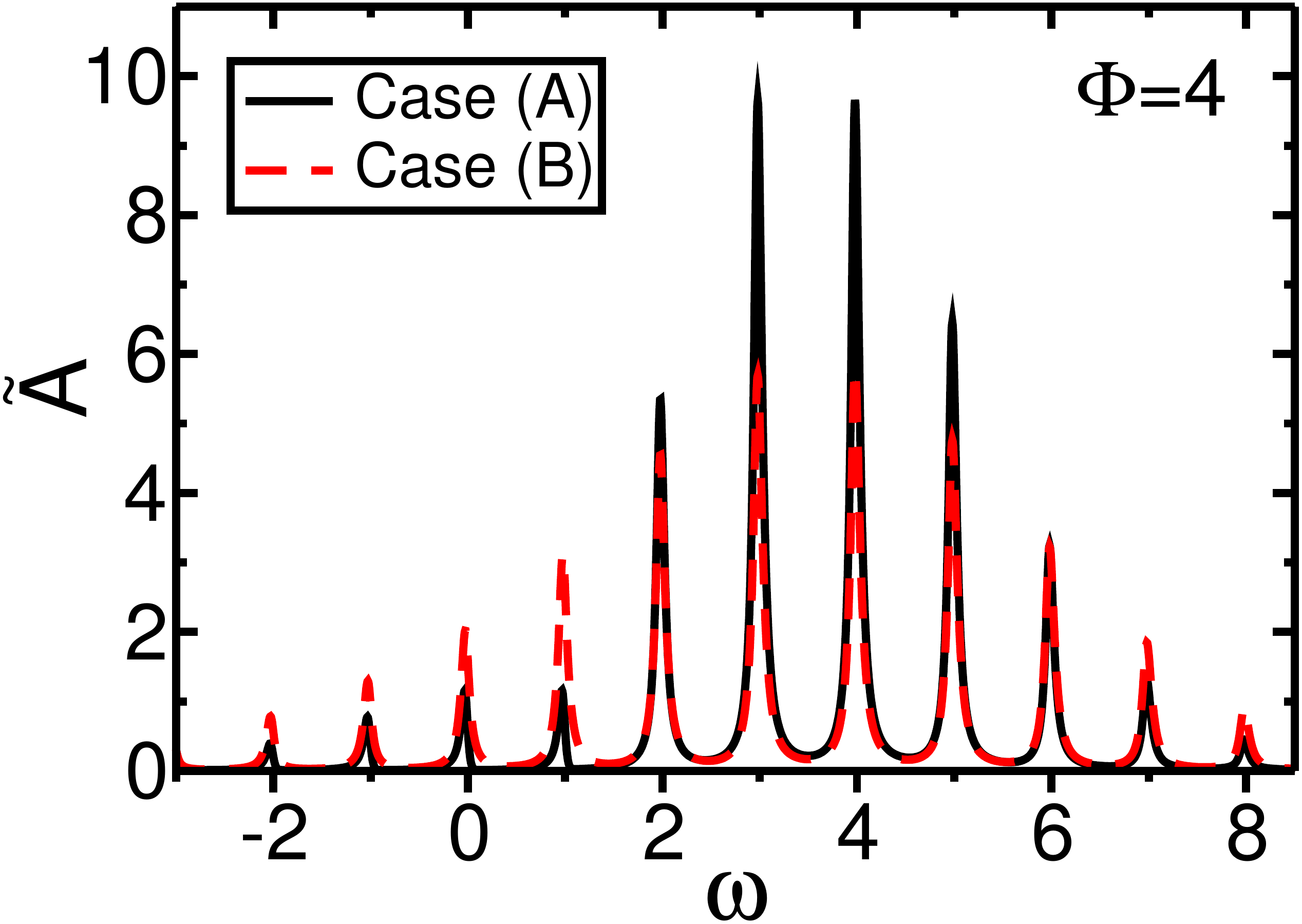}\quad 
\includegraphics[width=0.44\linewidth]{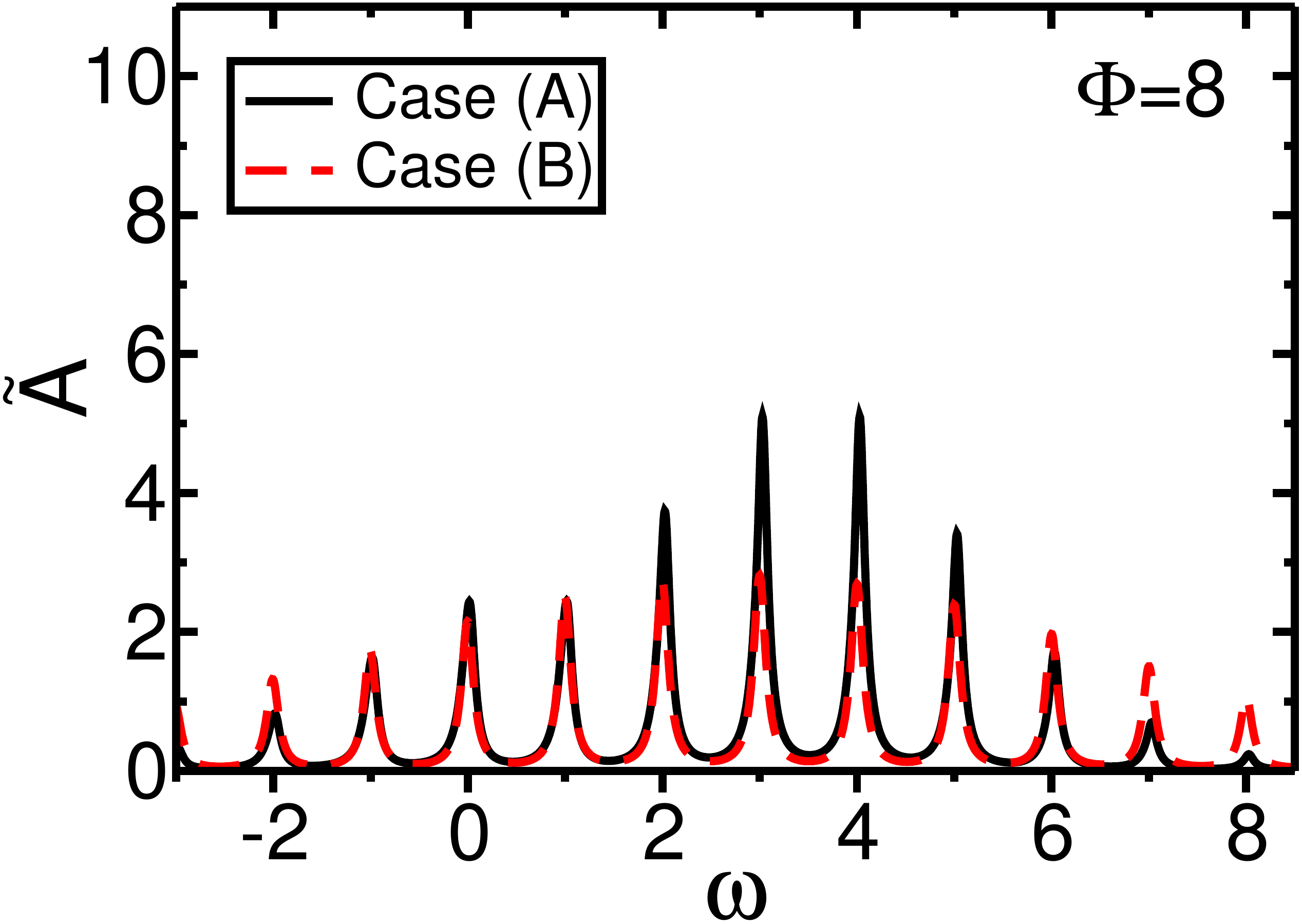}
\caption{Electronic spectral functions $\widetilde A(\omega)$ for cases (A)  and (B) at $\Phi=4$ (left) and   $\Phi=8$ (right), where $\varepsilon_p=2$. Again $\Gamma_0=0.1$, $\Delta=4$, $T_R=0.02$, and $\delta T=0$. }
\label{FIG3}
\end{figure}

At first, the response to the bias voltage $\Phi$ for $T_L=T_R$ and various  polaron energies $\varepsilon_p=g^2\omega_0$ is dealt with. Figure~\ref{FIG2} (upper row) reveals pronounced steps in the particle-current dependences $J(\Phi)$. These characteristic features  are manifestations of the well-separated multi-phonon resonances in the electronic spectral function displayed in Fig.~\ref{FIG3}. This becomes evident from the representation of $J$ by Eq.~\eqref{EQUcurrent}: The ``Fermi window'' given by  the difference $f_L(\omega+U_L)-f_R(\omega+U_R)$ is broadened with increasing $\Phi$ and, in this way, additional local maxima of $\widetilde A(\omega,U)$ contribute to $J$. Since the total dissipation rate (Fig.~\ref{FIG2}, second row) is proportional to $J$ its steps also occur in the heat flow to the phonon bath [given for case (A) in absolute values]  and the  effective temperature $T_P^{eff}$ (see Fig.~\ref{FIG2}, third row), as well as in the voltage dependences of the heat flows (see Fig.~\ref{FIG2}, fourth and fifth rows).  The dependence of the effective-temperature  on $\Phi$ shows the correct behavior for low voltages and exhibits the same
characteristic features as the corresponding quantities in~\cite{Galperin2007b}.  According to Fig.~\ref{FIG2}, fourth and fifth row, the heat currents into the system from the baths of the leads are negative, in agreement with the expectation that the heat generated by the energy dissipation is transfered to the heat baths.  Obviously, according to~\eqref{a16} and \eqref{EQUgelec}, a change of $\varepsilon_p$ causes a shift of the electronic spectral function along the $\omega$-axis  leading to the shift of the curves in Fig.~\ref{FIG2} observed for low $\Phi$-values. For high voltages, due to the Condon-blockade effect, the current response decreases with increasing $\varepsilon_p$. The smaller current response in case (B) as compared to case (A) may be attributed to the transfer of spectral weight to higher frequencies in case (B), compare Fig.~\ref{FIG3}.

\begin{figure}[t]
\hspace{-0cm}\includegraphics[width=0.32\linewidth]{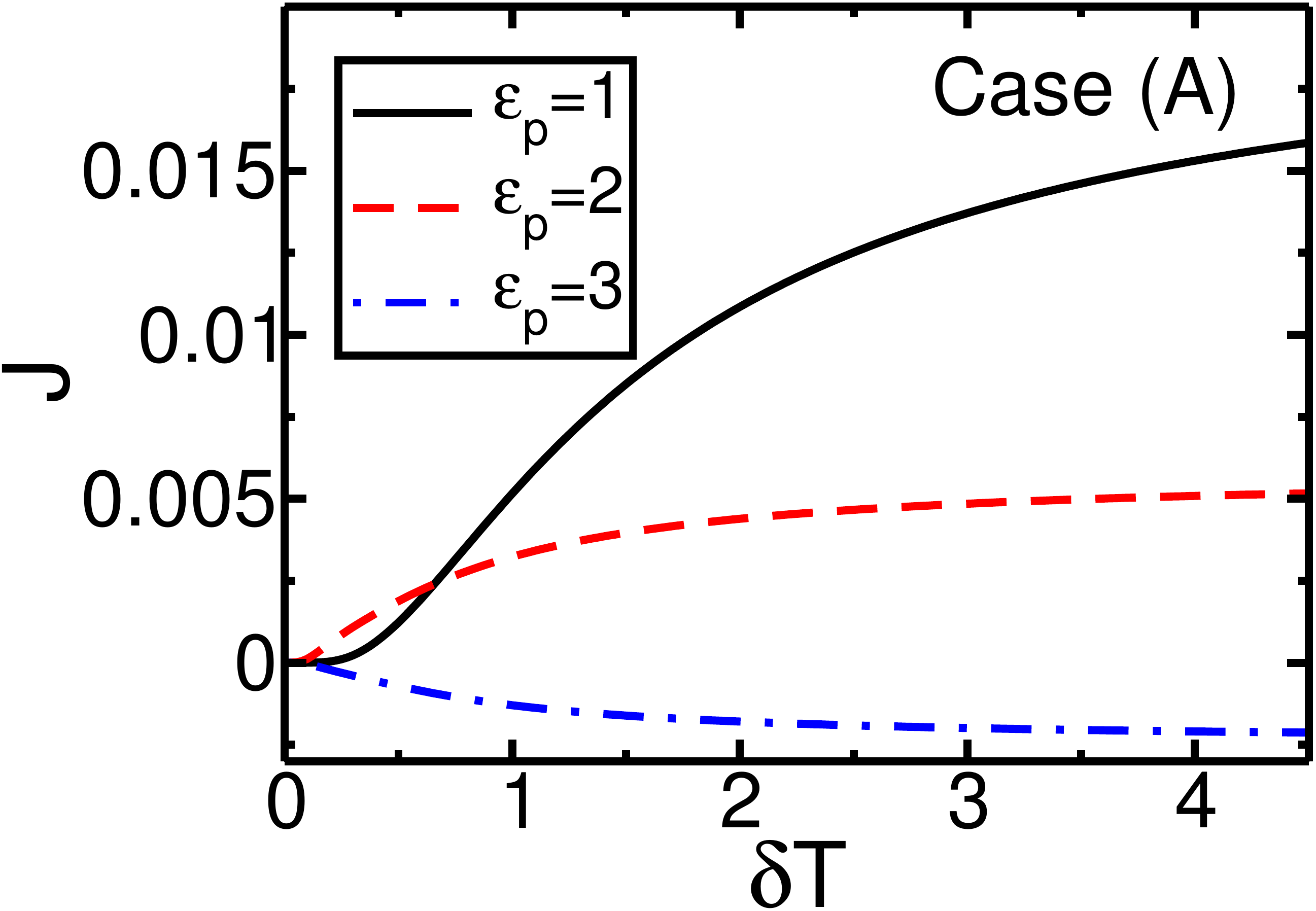}
\includegraphics[width=0.32\linewidth]{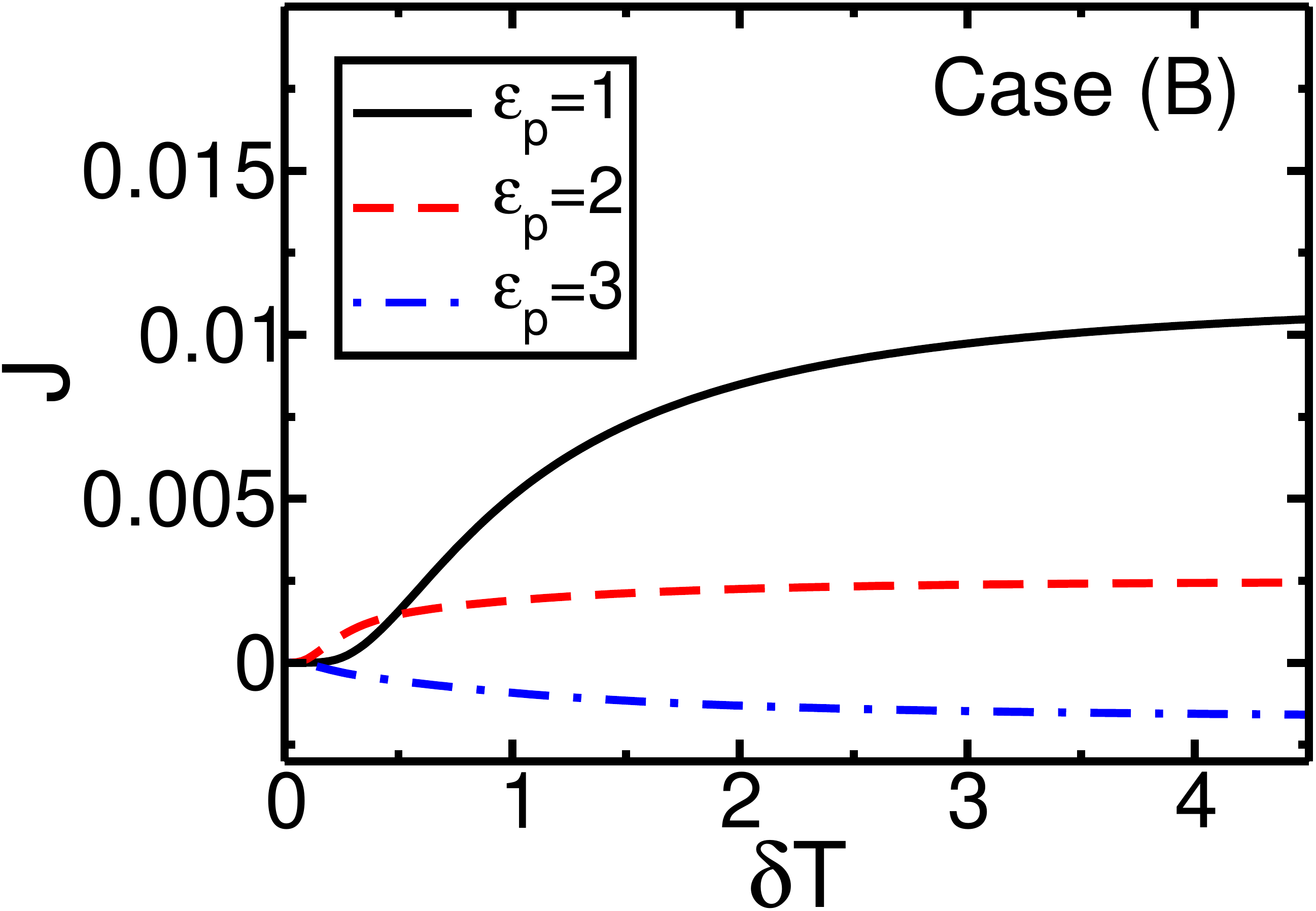}
\includegraphics[width=0.32\linewidth]{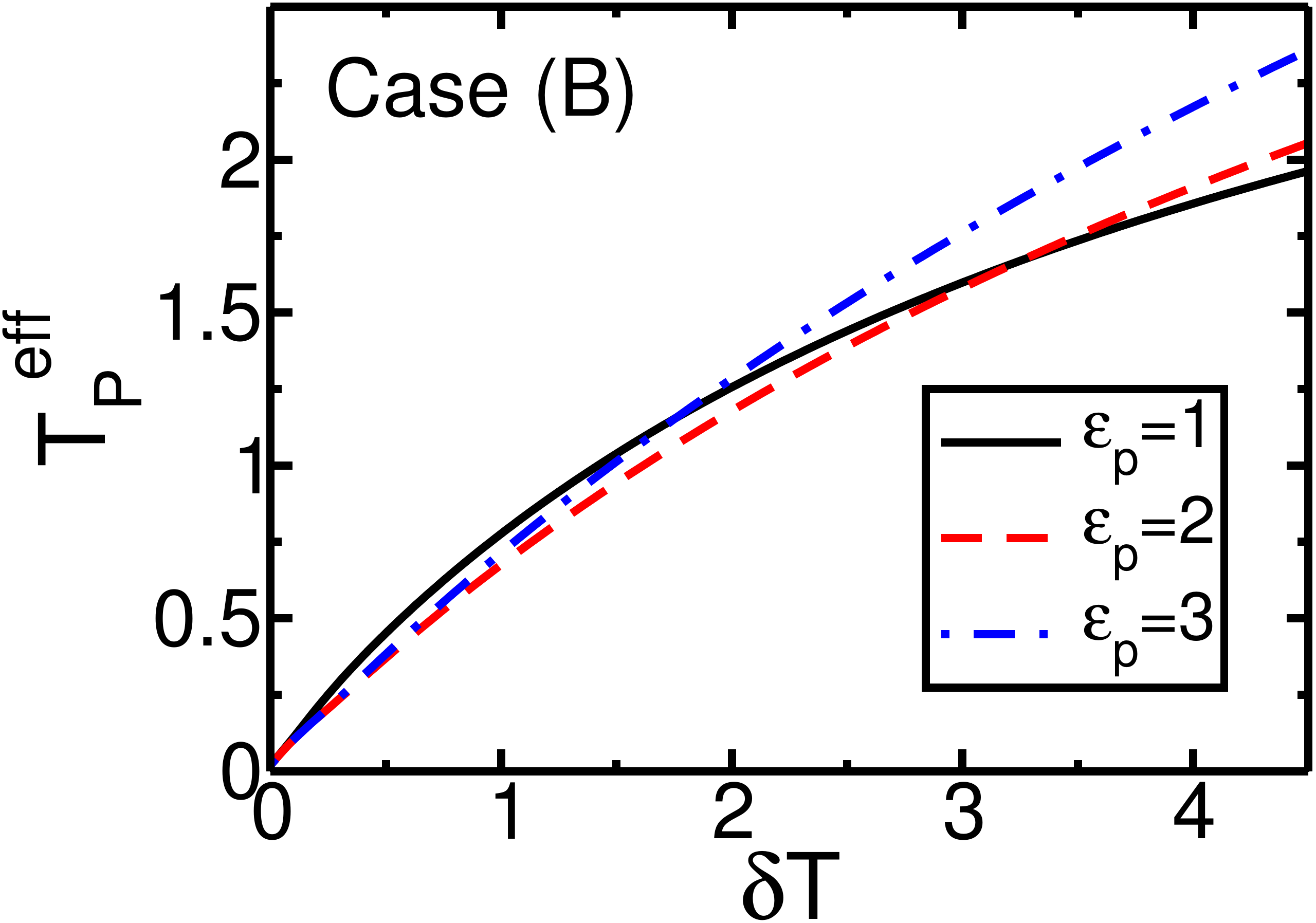}\\[0.1cm]
\includegraphics[width=0.32\linewidth]{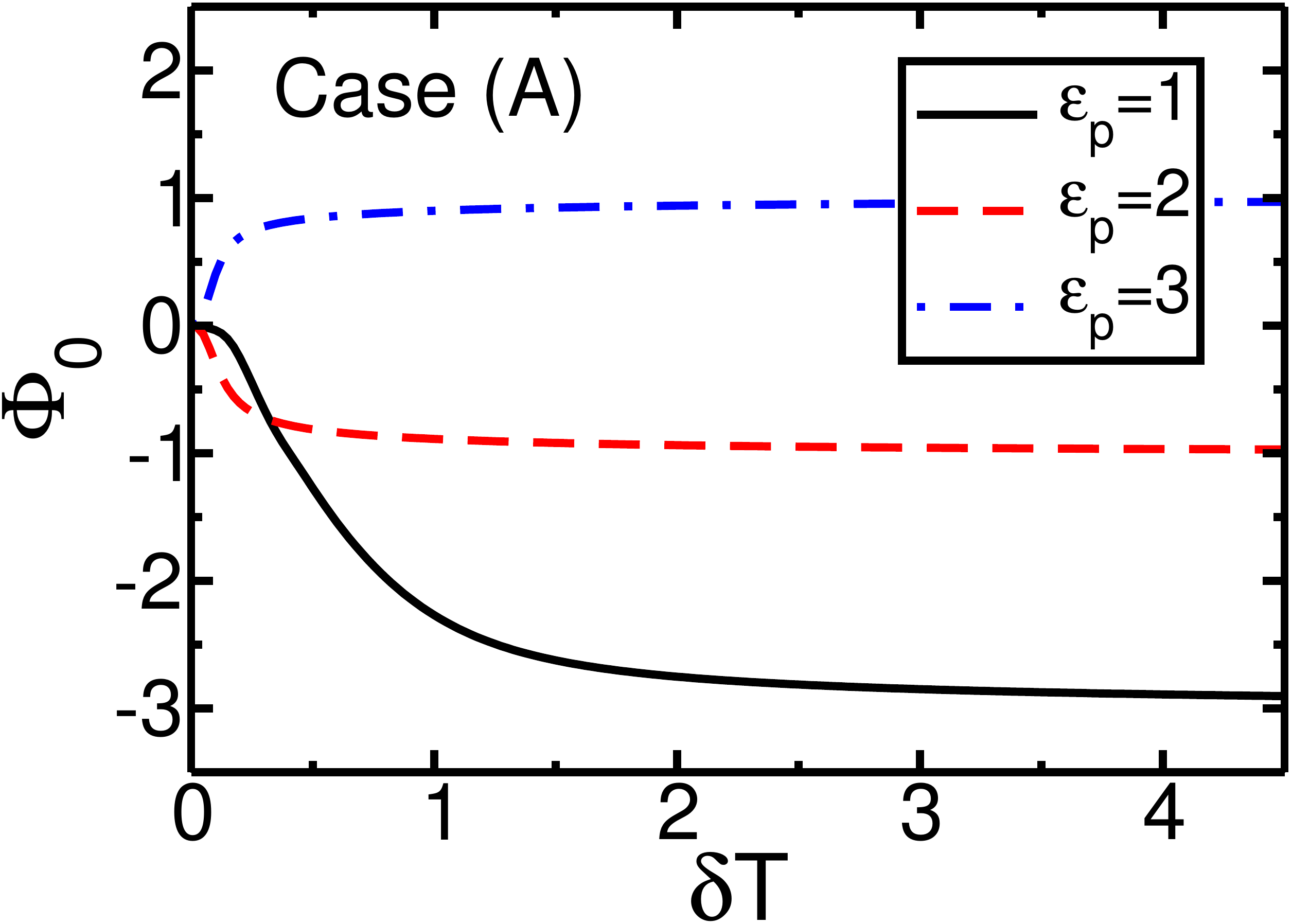}
\includegraphics[width=0.32\linewidth]{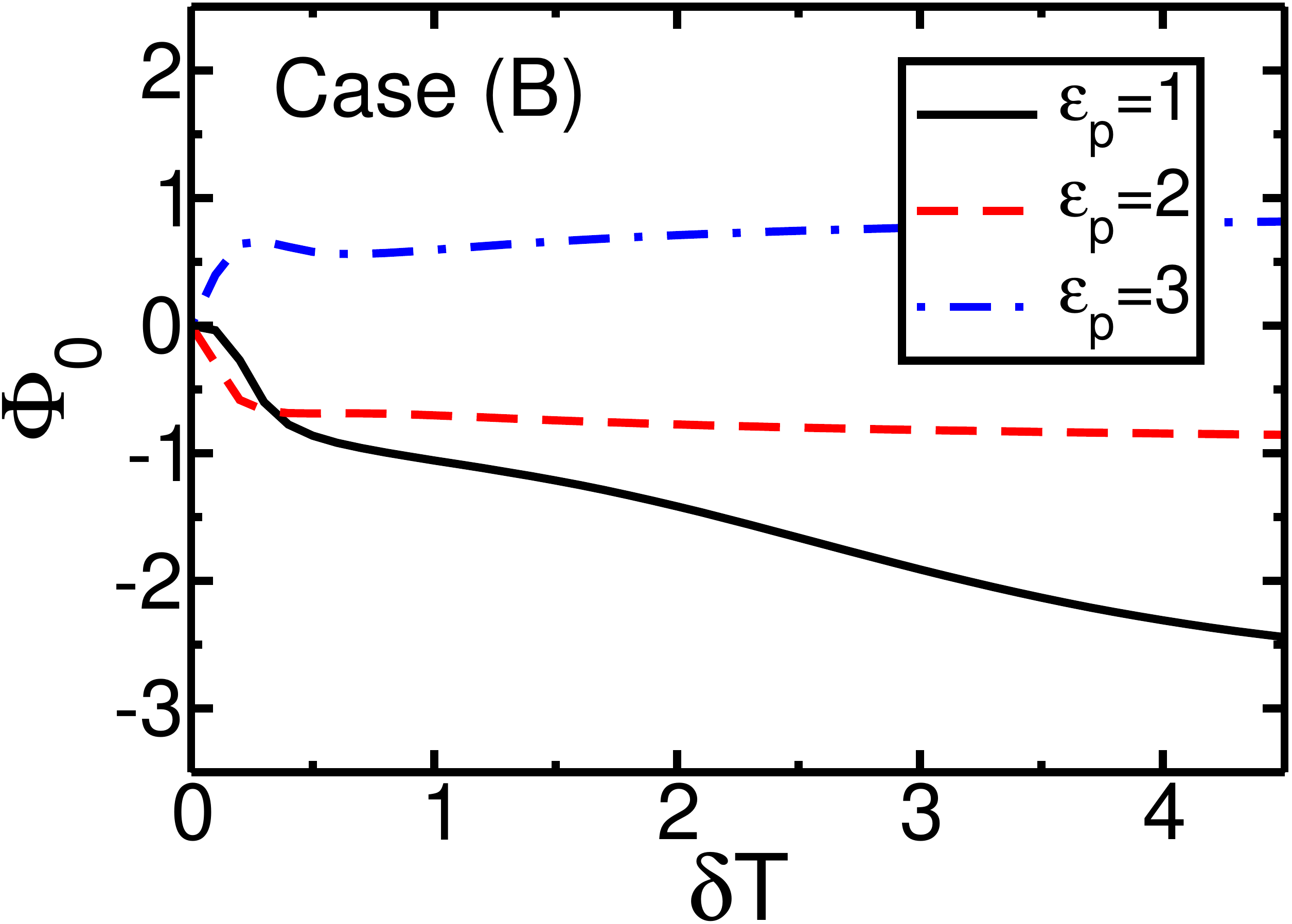}
\includegraphics[width=0.32\linewidth]{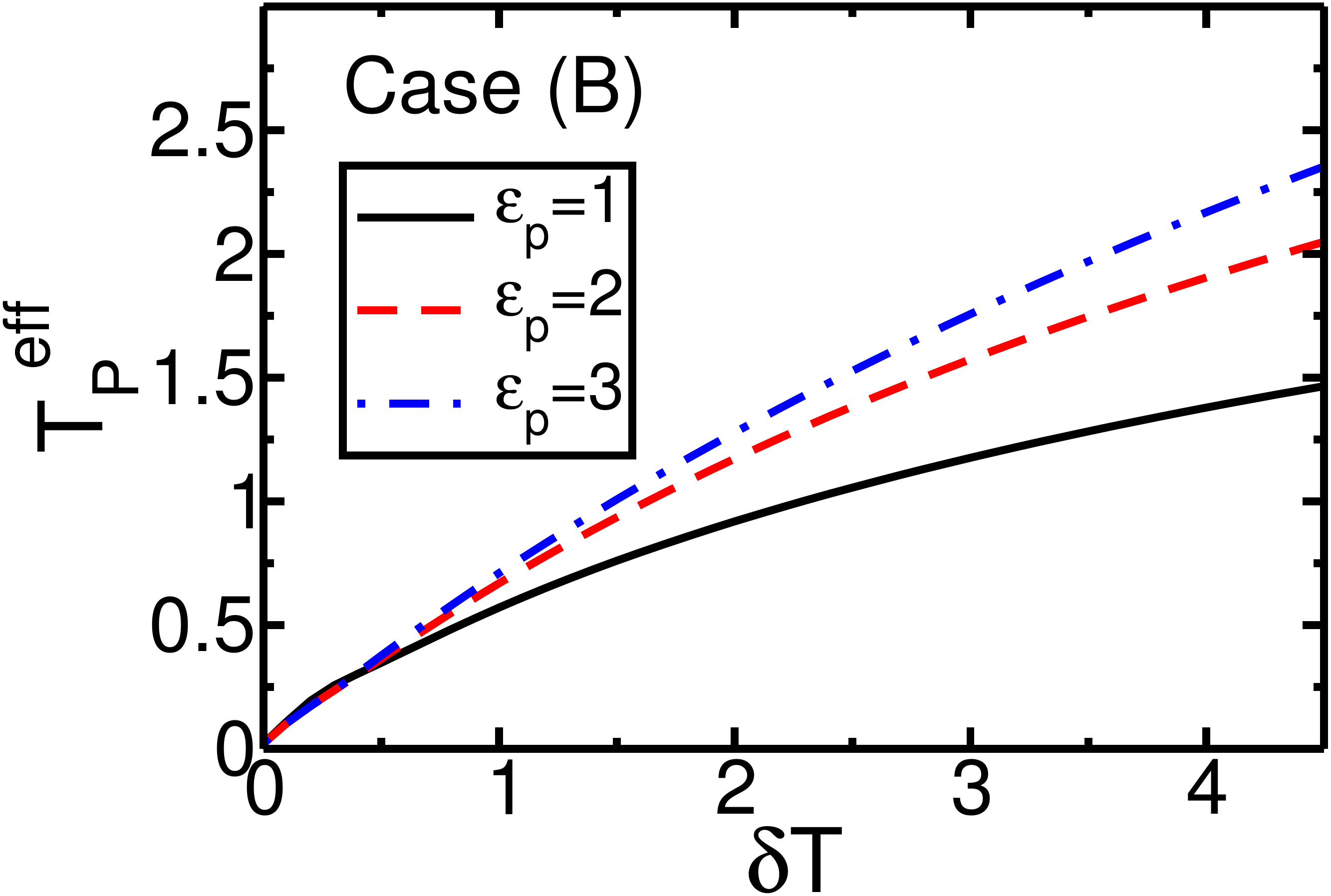}
\caption{Top: Current through the junction $J$ [cases (A) and (B)] and effective temperature $T_P^{eff}$  [case (B)]  as functions of the temperature difference $\delta T$ for $\Phi=0$. Bottom: Thermovoltage $\Phi_0$ [cases (A) and (B)] and effective temperature $T_P^{eff}$  [case (B)]  as functions $\delta T$. Other system parameters: $\Gamma_0=0.1$, $\Delta=2.5$, 
and $T_R=0.02$.}
\label{FIG4}
\end{figure}

\begin{figure}[t]
\includegraphics[width=0.44\linewidth]{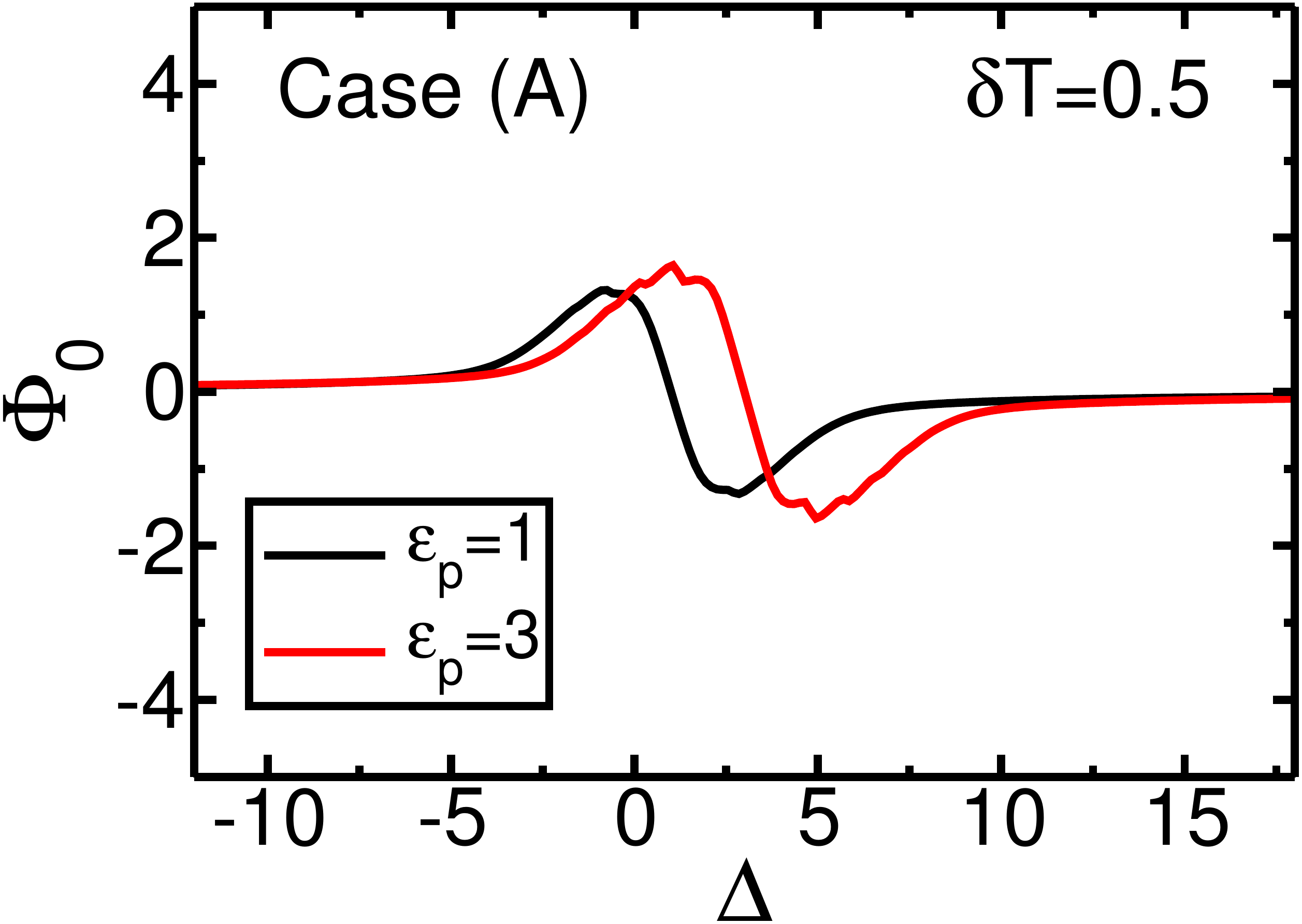}
\includegraphics[width=0.44\linewidth]{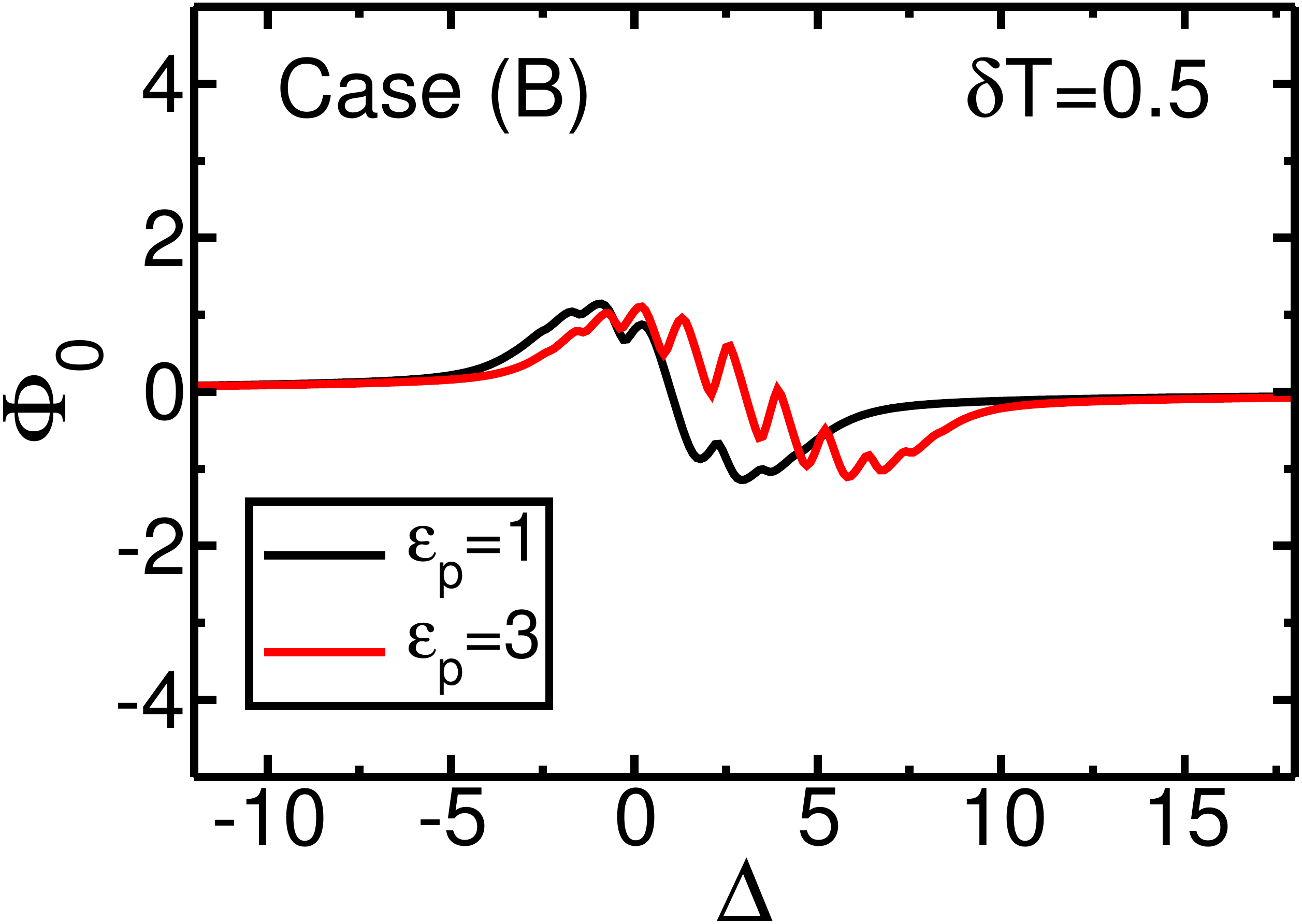}\\[0.2cm]
\includegraphics[width=0.44\linewidth]{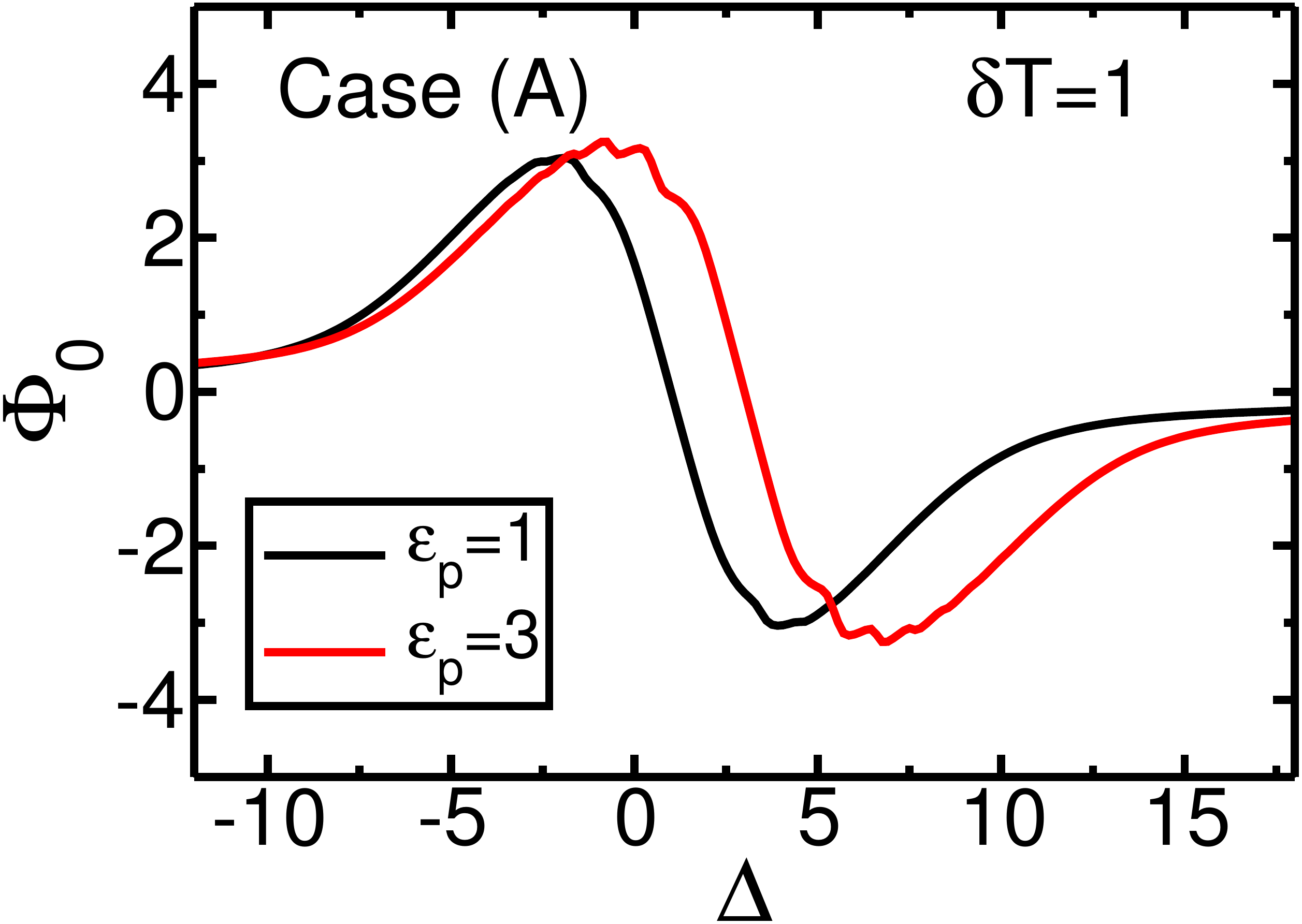}
\includegraphics[width=0.44\linewidth]{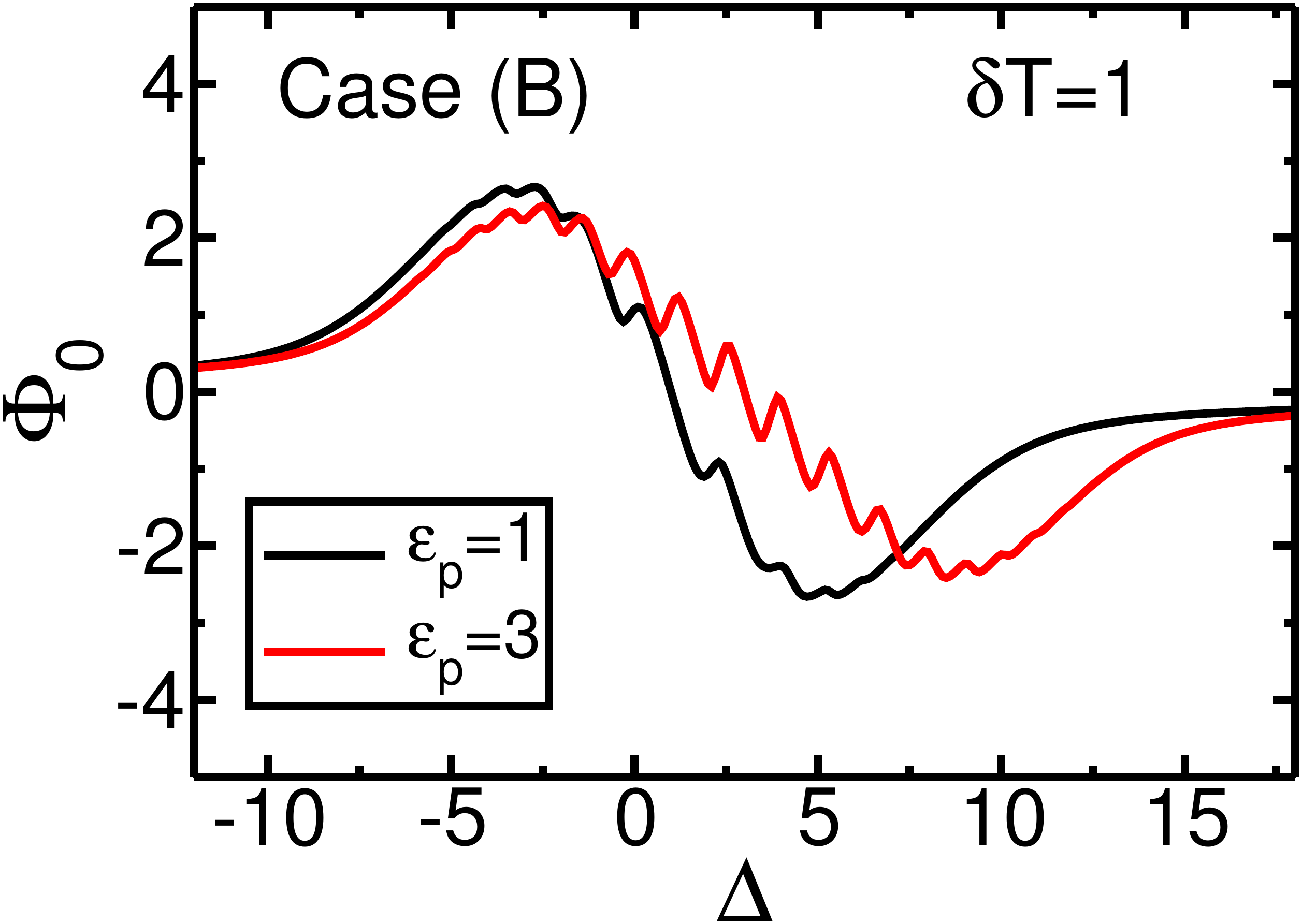}
\caption{Thermovoltage $\Phi_0$ in dependence on the dot-level energy $\Delta$ for cases (A) (left) and (B) (right) 
at $\delta T=0.5$ (top) respectively $\delta T=1$ (bottom), where $\Gamma_0=0.1$ and $T_R=0.02$.}
\label{FIG5}
\end{figure}

Taking in the formula for the particle transport~\eqref{EQUcurrent} the limit $\Phi=0$ at constant $\delta T= T_L-T_R$, the current response to the temperature difference between the leads is obtained, see Fig.~\ref{FIG4} (upper panels). By contrast, choosing the bias voltage such that $J=0$ at fixed $\delta T$, the quantity $\Phi_0$ known as thermovoltage is obtained (lower panels). The shape of the curves $J(\delta T)$ depends on the manner how the function $f_L(\omega)-f_R(\omega)$ overspreads the spectral function $\widetilde A(\omega)$ for $U_a=0$. Again the change of $\varepsilon_p$ results in a shift of the spectral function on the $\omega$-axis and, in particular, a sign-change of $J(\delta T)$ and $\Phi_0(\delta T)$ might happen, as it is demonstrated by the upper and lower panels of Fig.~\ref{FIG4}, respectively. The same effect is achieved by changing the unrenormalized dot energy level $\Delta$, too. The possibility to tune the thermovoltage by varying  the dot energy-level is illustrated in Fig.~\ref{FIG5}. 

\section{Conclusions}
To summarize, in this paper, we presented a nonequilibrium Green functions  approach that captures both the flow of heat and charge through a nanoscale junction in the steady-state regime. Thereby the leads assumed to be in local equilibrium, and maybe held at different temperatures and electric (chemical) potentials in order to drive the heat and particle currents, while the quantum dot placed between them is affected by molecular vibrations with a fixed phonon temperature (an effective temperature mainly determined by the EP interaction) in case the coupling to the phonon-heat bath is strong (rather negligible). Analyzing the thermoelectric transport properties of such a setup by  calculating, e.g., the heat/charge currents, the dot spectral function, the thermovoltage,  and the dot effective temperature in dependence of the voltage bias, the temperature difference between the leads, the energy of the dot level, and the electron-phonon coupling strength, we address several theoretical issues and fundamental open questions, such as how one defines  and what determines the local temperature (profile) along the system in a nonequilibrium situation? In doing so, we investigate the steady-state conditions. We furthermore demonstrate that polarons will play an important role in molecular thermoelectrics especially in the quantum regime, e.g., with regard to the Franck-Condon blockade. So far we have neglected any Coulomb interaction effects~\cite{FWLB08,HLCCSW13}; without doubt their inclusion would be a worthwhile future project.  
\section*{Acknowledgments}
This work was supported by Deutsche Forschungsgemeinschaft through SFB 652 B5. H.F. thanked the Institute of Physics of the ASCR for generous hospitality.

\begin{appendix}
\section{Polaron representation}\label{AA}
The calculations to the main text are carried out supposing the EP interaction is essentially stronger than the interaction of the dot with the leads. In analogy to the small polaron theory, the representation given by the Lang-Firsov transformation~\cite{LF62r}
\begin{align}
\label{lft}
S_g&=\exp\{ g (b^{\dag}-b)d^\dag d\}
\end{align}
is used. In this way, 
$\widetilde H =S_g^\dag H S_g$, reads
\begin{align}
\widetilde H &= \widetilde\Delta d^\dag d^{\nag}   + \omega_0  b^{\dag} b \label{EQUtransH}\\
& +\sum_{k,a} \left(\varepsilon_{ka}-\mu_a\right)c_{ka}^{\dag} c_{ka}^{\nag}-\sum_{k,a} \left(  C_{ka}^{\nag} d^{\dag}c_{ka}^{\nag}+ C_{ka}^\dag c_{ka}^{\dag}d\right) \,,  \nonumber
\end{align}
where
\begin{align}
\widetilde\Delta&=\Delta-\varepsilon_p\,,\;\, C_{ka}=\frac{t_{ka}}{\sqrt{N}}\,\mathrm{e}^{- g (b^{\dag}-b)}\,,\;\, \varepsilon_p=g^2\omega_0\,.
\end{align}
The so-called polaron binding energy $\varepsilon_p$ can be taken as a measure of the EP coupling strength. In Eq.~\eqref{EQUtransH}, the operators $d$ and $b$ represent the polaronic dot-state and the shifted oscillator-state, respectively. Electron operators, henceforth denoted by $\widetilde d$, are in the new representation expressed in terms of the polaron operators by the relation
\begin{align}
\widetilde d&= \exp\{ g (b^{\dag}-b)\} d\,.\label{EQUtilded}
\end{align}
Using the customary decoupling procedure, the electronic Green functions are expressed by the polaronic ones:
\begin{align}\label{EQUgelec}
\widetilde g^{\lessgtr} (\omega;U)&=\mathrm{e}^{- g^2\coth\theta}\Big\{ I_{0}(\kappa)g^{\lessgtr}(\omega;U)\nonumber\\&+ \sum_{s\ge 1} I_{s}(\kappa)2\sinh(s\theta)\nonumber\\
&\times\Big( [1+n_B(s\omega_0)]g^{\lessgtr}(\omega\pm s\omega_0;U)\nonumber \\
&+ n_B(s\omega_0)g^{\lessgtr}(\omega\mp s\omega_0;U)  \Big ) \Big\}\,,
\end{align}
where we have defined
\begin{align}
\theta&= \frac{1}{2}\beta_P\,\omega_0\;,\quad \kappa=\frac{ g^2}{\sinh \theta}\,,
\end{align}
and
\begin{align}
I_s(\kappa)&=\sum_{m=0}^{\infty}\frac{\left(\kappa/2\right)^{s+2m}}{m!(s+m)!}\,.
\end{align}

\section{Nonequilibrium Green functions}\label{AB}
Within  the Kadanoff-Baym formalism~\cite{Kadanoff1962} the real-time response functions are deduced using the equation of motion for the nonequilibrium Green functions 
of the complex-time variable $t=t_0-\mathrm{i}\tau$, $\tau\in[0,\sigma]$. In particular, 
\begin{align}
G_{dd}(t_1,t_2;U,t_0)&=-\frac{\mathrm{i}}{\langle S \rangle}\langle \mathcal{T}_\tau d(t_1)d^\dagger(t_2)S\rangle \label{EQUdefG}\;,\\
S=\mathcal{T}_\tau & \exp\Big\{-\mathrm{i}\int_{t_0}^{t_0-\mathrm{i}\sigma}\mathrm{d}t \;H_{U}(t)\Big\} \,.\label{EQUexponential}
\end{align}
Here, the time dependence of the operators is given by the unperturbed part of $H$, while the external disturbance is explicitly in the time-ordered exponential operator $S$. The operator $T_\tau$ orders the operators by the imaginary parts of the times. The function $G_{dd}$ is equal to the analytical functions $G_{dd}^>(t_1,t_2;U,t_0)$ and $G_{dd}^<(t_1,t_2;U,t_0)$  for $\mathrm{i}(t_1-t_2)>0$ and $\mathrm{i}(t_1-t_2)<0$, respectively. In comparison with the generalized temperature Green functions defined in Ref.~\cite{Kadanoff1962}, where $\sigma=\beta=(k_BT)^{-1}$ is taken, we make a generalization by assuming the value of $\sigma$ to have no specific physical meaning~\cite{Koch2014}. The latter assumption is to be made because the temperature may not be homogeneous throughout the system in the steady states considered. Therefore, the function defined by ~\eqref{EQUdefG}, \eqref{EQUexponential} does not have the properties of the temperature $\beta$-dependent Green functions; it rather represents a functional of the ordered operators which is used to determine the real-time response function. In particular,
\begin{align}
\lim_{t_0\to-\infty} G_{dd}^{\lessgtr}(t_1,t_2;U,t_0) &= g_{dd}^{\lessgtr}(t_1,t_2;U)\label{EQUlimitlessgtr}\,,
\end{align}
and similar relations hold for $G_{cd}^{\lessgtr}$ and $g_{cd}^{\lessgtr}$.

Evaluating the real-time Green functions needed for the  calculation of  the stationary particle current through the molecular junction, we start with the equation of motion for the function $G_{cd}$, i.e.,
\begin{align}
G_{cd} & (k,a;t_1,t_2;U,t_0) = \nonumber\\
&-\frac{ t_{ka}^\ast}{\sqrt{N}} G_{cc}^{(0)}(k,a;t_1,\bar t;U) \bullet G_{dd} (\bar t,t_2;U,t_0)\,,\label{EQUcomplexeom1}
\end{align}
where $G_{cc}^{(0)}$ are the Green functions of the leads which are assumed to be in the local equilibrium state determined by $\mu_a$ and $T_a$. In Eg.~\eqref{EQUcomplexeom1}, the matrix multiplication ``$\bullet$'' denotes the integration $\int_{t_0}^{t_0-\mathrm{i}\sigma}\cdots\mathrm{d}\bar t$ over the complex arguments. Performing the limit $t_0\to-\infty$ while keeping $\mathrm{i}(t_1-t_2)<0$, the expressions of $g_{cd}^<(k,a;t_1,t_1;U)$ by means of $g^{(0)\lessgtr}_{cc}$ and $g_{dd}^{\lessgtr}$ is obtained. Then, using the formal manipulations explained in~\cite{Koch2011}, the equation~\eqref{EQUcurrentJa} results.

\section{Steady-state solution}\label{AC}
Starting from the Dyson equation for the polaron Green function of complex time,
\begin{align}
\left[G_{dd}^{(0)-1}(t_1,\bar t)-\Sigma (t_1,\bar t;U,t_0)\right]\bullet\,& G_{dd}(\bar t,t_2;U,t_0)\;\nonumber \\
&=\delta(t_1-t_2)\;,\label{EQUdyson}
\end{align}
with the inverse zeroth-order Green function
\begin{align}
G_{dd}^{(0)-1}(t_1,t_2)=\Big( \mathrm{i}\frac{\partial}{\partial t_1} - \widetilde \Delta \Big) \delta (t_1-t_2)\,,\label{EQUinverseG0}
\end{align}
the exact steady-state equations for the Fourier transforms of the real-time Green functions were deduced in~\cite{Koch2011}. The solution of theses equations was shown to have the form analogous to the corresponding equilibrium expressions, namely (with indices ``$d$'' omitted): 
\begin{align}
g^<(\omega,U)&=A(\omega,U)\bar f(\omega;U)\,,\\
 g^>(\omega,U)&=A(\omega,U)(1-\bar f(\omega,U))\;,\\
 A(\omega,U)&=\frac{\Gamma(\omega,U )}{\left[\omega-\widetilde\Delta-\mathcal{P}\int\frac{\mathrm{d}\omega^\prime}{2\pi}\;\frac{\Gamma(\omega^\prime,U )}{\omega-\omega^\prime}\right]^2+\left[\frac{\Gamma(\omega,U )}{2}\right]^2}\,,\label{a16}\\
 \Gamma(\omega,U )&=\Sigma^>(\omega,U )+\Sigma^<(\omega,U )\,,\\
\bar f (\omega,U )&=\frac{ \Sigma^<(\omega,U )}{ \Gamma(\omega,U )}\,.
\end{align}
Consequently, the quantities being crucial to make the above solution explicit are the polaron self-energy functions $\Sigma(\omega,U)$, which were evaluated in\cite{Koch2014,Koch2011} to second order in the coupling operators $C_{ka}^{(\dag)}$ with the following result:
\begin{align}
\Sigma^{<}&(\omega;U )=  \mathrm{e}^{-g^2\coth\theta} \sum_a \Big \{ I_0(\kappa) \Gamma^{(0)}(\omega)f_a(\omega+U_a)\nonumber\\
&+\sum_{s\ge 1} I_{s}(\kappa) 2 \sinh (s \theta) \nonumber\\
&\times\Big [ \Gamma^{(0)}(\omega-s\omega_0)n_B(s \omega_0) f_a(\omega-s\omega_0+U_a) \nonumber\\
&+\Gamma^{(0)}(\omega+s\omega_0) \big[n_B(s\omega_0)+1\big] f_a(\omega+s\omega_0+U_a)  \Big ] \Big \} \,,\label{EQUSigmalessfourier1}
\end{align}
where $\Sigma^{>}(\omega,U)$ is obtained from~\eqref{EQUSigmalessfourier1} by interchanging $n_B$ with $(n_B+1)$  and $f_a$ with $(1-f_a)$. 

\end{appendix}

%\bibliographystyle{unsrt}
%\bibliographystyle{plain}
%\bibliographystyle{apsrev4-1}
%\bibliography{ref}
%\bibliography{./thermo}
\end{document}